%% file: main_first_revision_submission.tex
\documentclass[journal,romanappendices]{IEEEtran}
\ifCLASSINFOpdf
\else
\fi
\usepackage{amsmath}
\usepackage{array}

\ifCLASSOPTIONcompsoc
  \usepackage[caption=false,font=normalsize,labelfont=sf,textfont=sf]{subfig}
\else
  \usepackage[caption=false,font=footnotesize]{subfig}
\fi
\usepackage{cite}
\usepackage{amssymb,epsfig}
\usepackage{graphicx}
\usepackage{color}
\usepackage[mathscr]{eucal}
\usepackage{tikz}
\usepackage{verbatim}
\usepackage{multicol}
\usepackage{array}
\usepackage{float}
\usepackage{amsbsy}
\usepackage{bm}
\usepackage{array}
\usepackage{multirow}
\usepackage{soul,xcolor}
\usetikzlibrary{shapes,arrows}
\usepackage{amsmath,bm,times}
\usepackage{soul,xcolor}
\usepackage{bm}
\usepackage{algorithm,algcompatible}
\usepackage{amsthm}
\usepackage{enumerate}
\usepackage{arydshln}
\usepackage{kbordermatrix}
\usepackage{amsfonts}
\usepackage{mathtools}
\usepackage{makecell}
\usepackage[hidelinks]{hyperref}

\theoremstyle{definition}
\newtheorem{definition}{Definition}[section]
\newtheorem{theorem}{Theorem}[section]
\newtheorem{proposition}{Proposition}[section]
\newtheorem{lemma}{Lemma}[section]

\pgfdeclarelayer{background}
\pgfdeclarelayer{foreground}
\pgfsetlayers{background,main,foreground}
\tikzstyle{rectanglebox}=[draw, fill=blue!20, text width=4em, 
    text centered, minimum height=4em]
\tikzstyle{ann} = [above, text width=5em]
\tikzstyle{naveqs} = [rectanglebox, text width=5em, fill=red!20, 
    minimum height=2.5em, rounded corners]

\usetikzlibrary{shapes,arrows,positioning}
\input{macros.tex}
\newsavebox{\smlmat}
\savebox{\smlmat}{$\left(\begin{smallmatrix}2&-2\\-1&2\end{smallmatrix}\right)$}

\algnewcommand{\Inputs}[1]{%
  \Init \textbf{Initialization:}
  \Initial \hspace*{\algorithmicindent}\parbox[t]{.8\linewidth}{\raggedright #1}
}

\setstcolor{red}

\begin{document}
%
\title{On PMU Data Integrity under GPS Spoofing Attacks: A Sparse Error Correction Framework}

%
%

\author{Shashini~De~Silva,
        Jinsub~Kim,~\IEEEmembership{Member,~IEEE,}
        Eduardo~Cotilla-Sanchez,~\IEEEmembership{Senior Member,~IEEE}
        and Travis~Hagan
\thanks{This material is based upon work supported by the Department of Energy under Award Number DE-OE0000780.}
\thanks{Part of this work was presented at the IEEE Global Conference on Signal and Information Processing, Anaheim, CA, 2018~\cite{Desilva_sparseerrorcorrection_2018}.}}

\maketitle

\begin{abstract}
Consider the problem of mitigating the impact on data integrity of phasor measurement units (PMUs) given a GPS spoofing attack. We present a sparse error correction framework to treat PMU measurements that are potentially corrupted due to a GPS spoofing attack. We exploit the sparse nature of a GPS spoofing attack, which is that only a small fraction of PMUs are affected by the attack. We first present attack identifiability conditions (in terms of network topology, PMU locations, and the number of spoofed PMUs) under which data manipulation by the spoofing attack is identifiable. The identifiability conditions have important implications on how the locations of PMUs affect their resilience to GPS spoofing attacks. To effectively correct spoofed PMU data, we present a sparse error correction approach wherein computation tasks are decomposed into smaller zones to ensure scalability. We present experimental results obtained from numerical simulations with the IEEE RTS-96 and IEEE 300 test networks to demonstrate the effectiveness of the proposed approach.  
\end{abstract}


\section{Introduction}
%
%
%
%
\IEEEPARstart{P}{hasor} measurement units (PMUs), which are equipped with clocks synchronized by global positioning systems (GPS), or, more broadly, global navigation satellite systems (GNSS), provide direct measurements of voltage and current phasors at a much faster rate than the legacy SCADA system~\cite{NASPI_Synch_Data}. Due to this enriched measurement quality, there has been a wide interest in developing approaches to leverage PMU measurements for real-time power grid monitoring, protection, and control~\cite{de2010synchronized, NASPI_Synch_Mon}. While several PMU-based approaches showed improved performance compared to the legacy approaches~\cite{lassetter2017learning, brahma2016real}, those promises can be realized only if the data integrity of PMUs can be ensured.

Compared to legacy measurement devices, PMUs are equipped with more sophisticated security protocols, and thus it is considered difficult for adversaries to tamper with the data by directly compromising data authentication protocols of PMUs~\cite{bobba2010exploring}. Nevertheless, the dependency of PMUs on civilian GPS signals for clock synchronization renders PMU measurements vulnerable to GPS signal spoofing attacks, which can be successfully launched by an adversary with small resource demand. The cyber attackers can easily deploy GPS transmitters to broadcast counterfeit GPS signals, which can manipulate the time estimation at the target PMU's GPS receiver~\cite{shepard2012evaluation,jiang2013spoofing}. Erroneous time reference successively induces errors in phase angle measurements of the tampered PMU. In practice, an attacker with limited resources can spoof only a few PMUs at a time. Hence the impact of spoofing attacks on PMU measurements is sparse in nature,~\ie, we can assume the fraction of PMU measurements corrupted by the spoofing attacks to be small.

In this paper, we exploit the sparse nature of the GPS spoofing attack to recover affected PMU measurements. 
We first derive identifiability conditions for GPS spoofing attacks, under which a spoofing attack is fundamentally identifiable. This identifiability assessment portrays the vulnerability of the PMU network to GPS spoofing attacks and can be used to determine PMU placement that is resilient to GPS spoofing attacks. Then, we develop a sparse error correction algorithm to effectively correct potentially spoofed PMU measurements. The decomposability of the PMU measurement model is leveraged to make major steps of the algorithm performed based on only local models and measurements thereby ensuring the scalability of the algorithm. The experiment results on IEEE RTS-96 and IEEE 300 bus test networks show that the proposed approach outperforms other benchmark algorithms when the number of spoofed PMUs is moderate.

\subsection{Related work}
There have been a plethora of work conducted on false data injection attacks and state estimation using conventional sensor measurements from power grid. The scope of this work range from designing undetectable sparse data attacks that can manipulate state estimation solutions~\cite{Kosut_malicious_data_attacks_2011,Ozay_sparse_attack_2013,Teixiera_cyber_security_2010} to developing robust state estimation techniques~\cite{Hao_sparse_malicious_2015,Jin_power_grid_ac_based_2019,Teixeira_secure_control_2015,Weiyu_sparse_error_correction_2013}. The authors in~\cite{Zhang_false_data_injection2017,alexopoulos2020complementarity} have studied false data injection attacks on PMU measurements and their impact on PMU-based state estimation. In the above works, a common assumption on false data injection attack is that the adversary is capable of manipulating the measurements from the compromised sensors to \emph{any desirable values}. False data correction strategies developed for this adversary model can be applied for correcting generic false data entries~\cite{Weiyu_sparse_error_correction_2013}. However, they can be suboptimal in mitigating the impact of GPS spoofing attacks on PMU measurements because these approaches do not take into account the unique characteristics of how GPS spoofing attacks affect the measurements from the spoofed PMUs. In what follows, we discuss existing literature that defend and correct GPS spoofing attacks on PMU measurements.

Several approaches have been proposed in the literature to enhance the resilience of GPS time estimation procedure at a \emph{single} PMU against spoofing attacks by developing a new GPS receiver architecture~\cite{gong2012gps, heng2014reliable,yu2014short} or a new robust time estimation algorithm~\cite{ng2016robust,bhamidipati2016multi}. Gong \emph{et al.} in~\cite{gong2012gps} proposed a spoofing detection algorithm by leveraging multiple GPS receivers per PMU, while authors in~\cite{heng2014reliable} exploit the networked and static nature of PMUs in close proximity, to propose a robust receiver architecture. Also in~\cite{yu2014short} the authors leverage the characteristics of a static receiver network to constrain the adversary's freedom of GPS spoofing. The main focus of all the aforementioned techniques is on designing robust receiver architectures that harden the spoofing attacks. On the other hand, authors in~\cite{ng2016robust,bhamidipati2016multi} propose robust time estimation techniques such that time estimation in PMUs become resilient to spoofing attacks. Authors in~\cite{ng2016robust} propose a direct time estimation technique using the maximum likelihood approach. Work in~\cite{bhamidipati2016multi} couples this time estimation technique with spatially dispersed multiple-receivers to improve the resilience against spoofing attacks. All the strategies in this category either require additional infrastructure in terms of external clocks and multiple GPS receiver antennas or require a network of GPS receivers in the vicinity of the PMU of interest. 

The aforementioned works focused on robustifying the time estimation procedure at a single PMU. In the meanwhile, several works in the literature~\cite{pradhan2016gps,fan2017synchrophasor,risbud2018vulnerability,RisbudMultiperiod,vanfretti2010phasor,ghiocel2013phasor} demonstrated that GPS spoofing attacks on PMUs can be more effectively mitigated by leveraging how phasor measurements from different PMUs are correlated and how they are related to the underlying power system state due to the interconnectedness of the grid. Pradhan \emph{et al.} in~\cite{pradhan2016gps} present a dynamic state estimation from PMU measurements that is resilient to spoofing attacks, by devising a generalized likelihood-based hypothesis testing to detect the location and magnitude of the spoofing attacks. A major limitation here is that it assumes that an accurate estimate of the time of the attack is known a priori. In~\cite{fan2017synchrophasor} the authors mathematically model the spoofed measurements and propose an algorithm to detect and correct GPS spoofing attack on a single PMU in the network. Risbud \emph{et al} in~\cite{risbud2018vulnerability, RisbudMultiperiod} leverage a measurement model that accounts for GPS spoofing attacks on multiple PMUs and develop an alternating minimization algorithm for joint estimation of the state and the phase angle biases in the PMU measurements introduced by the spoofing attacks. Similarly the authors of~\cite{vanfretti2010phasor,ghiocel2013phasor} attempt to jointly estimate the states and phase angle biases by solving weighted least squares problem. However, joint estimation of state and phase angle biases, without an additional assumption on the state or angle bias variables, is ill-posed in that there exist many distinct solutions that can fit PMU measurements optimally. 
\subsection{Summary of Contributions}
In this paper, we consider the PMU measurement model in the presence of GPS spoofing attack and formulate the problem of estimating phase angle biases introduced by the spoofing attack. In order to address the aforementioned ill-posedness issue, we impose a practical constraint on the spoofing attack that only a small fraction of PMUs are subject to spoofing attacks at a given time. Under this assumption, we develop and analyze a sparse error correction framework for estimating the \emph{sparse} phase angle biases introduced by the spoofing attack. Our formulation is intended for direct estimation of the phase angle biases (without a need to jointly estimate the state), and thus our approach does not require state observability based on PMU measurements;~\ie, the approach is applicable even when the state is not observable based on PMU measurements. 
The main contributions are as follows:
\begin{enumerate}[i)]
\item We formulate PMU data correction under GPS spoofing attacks as a sparse error correction problem. 
\item We present a rigorous identifiability analysis which provides simple conditions (in terms of the network topology, PMU locations, and the number of spoofed PMUs) under which  the phase angle biases introduced by spoofing attacks are fundamentally identifiable in the sparse error correction framework. These conditions provide an important insight on how the PMU locations affect the resilience of PMU measurements to GPS spoofing attacks. 
\item We present a scalable sparse error correction algorithm to estimate sparse phase angle biases introduced by the spoofing attacks and correct the PMU measurements.
\item We validate the approach using the MATLAB experiments with RTS-96 and IEEE 300-bus networks. The approach outperformed benchmarks~\cite{risbud2018vulnerability,vanfretti2010phasor} in correcting PMU measurements in the presence of GPS spoofing attacks on multiple PMUs, under both observable and unobservable PMU settings.
\end{enumerate}


\section{Problem formulation}
\label{sec:probform}
Throughout the paper, boldface lowercase letters (\textit{e.g.}, $\bf{x}$) denote vectors, boldface uppercase letters (\textit{e.g.}, $\bf{X}$) denote matrices and script letters (\textit{e.g.}, $\mathscr{X},\mathscr{A}$) denote sets. The symbols $\mathbb{R}$ and $\mathbb{C}$ are used to denote real number and complex number domains, respectively. For instance, an $n$-dimensional vector $\xbf$ in real domain is indicated by $\xbf\in\mathbb{R}^n$. The $l_p$ norm of $\bf{x}$ is denoted by $\|\xbf\|_p$. Furthermore, $|x|$, $\angle x$, and $x^*$ denote the magnitude, the angle, and the Hermitian transpose of the complex number $x$, respectively. Moreover, for a sparse vector $\xbf$, $supp(\xbf)$ denotes the support of $\xbf$, which is the set of indices of nonzero entries in $\xbf$. In addition, $\Rmsc(\Hbf)$ and $\Nmsc(\Hbf)$ denote the range space and the null space of $\Hbf$ respectively. 

\subsection{PMU measurement model}
A power network topology can be represented by an undirected graph $\mathscr{G} = (\mathscr{V}, \mathscr{E})$ where $\mathscr{V} = \{1, 2, ..., N\}$ denotes the set of $N$ buses in the network and $\mathscr{E}$ denotes the set of branches (either transmission lines or transformers) interconnecting these buses, specifically,~$\{i,l\}\in\mathscr{E}$ if and only if there exists an energized line connecting bus $i$ and bus $l$. PMUs are installed in a selected subset of buses $\mathscr{T} \subseteq \mathscr{V}$ where total number of PMUs in the network is denoted by $K$ (\ie,~$|\mathscr{T}| = K$). 


Let $\zbf\in\mathbb{C}^{m}$ denote the PMU measurement vector which consists of voltage and current phasor measurements from all the PMUs deployed in the network and $\xbf = [x_1, x_2, \dots, x_N]^T\in\mathbb{C}^{N}$ denote the complex system state vector where $x_i$ represents the voltage phasor at bus $i$, precisely $x_i = |x_i|e^{j\angle x_i}$. If a PMU is installed at bus $i$ (\ie, $i \in \Tmsc$), $\zbf$ would contain the measurement of the voltage phasor at bus $i$, which we denote by $z_{V_i}$:
\begin{equation}
\label{eq: voltage phasor}
\begin{aligned}
    z_{V_i} & =  x_i
\end{aligned}
\end{equation}
In addition, PMU at bus $i$ also provides measurements of outgoing current phasors in a subset of lines incident to bus $i$. Suppose $\mathscr{N}_i$ denotes the set of neighbours of bus $i$ in the topology $\Gmsc$, and $\Mmsc_i\subseteq\Nmsc_i$ denotes a subset of neighbors of bus $i$ such that the current phasor of line $\{i,l\}$ with $l\in\Mmsc_i$ is measured by PMU at bus $i$. The complex phasor measurement of the line current from bus $i$ to bus $l\in\Mmsc_i$, denoted by $z_{I_{il}}$, can be given as below:
\begin{equation}
\label{eq: current phasor}
   \begin{aligned}
  z_{I_{il}} =&~y_{il}(x_i - x_l) + j\frac{b^{s}_{il}}{2}x_i\\
    =&~(y_{il} + j\frac{b^{s}_{il}}{2})x_i - y_{il}x_l~, 
\end{aligned}
\end{equation}
where $y_{il}$ is the series admittance of the line $\{i,l\}$, and $b^{s}_{il}$ is its line charging susceptance. 

From~\eqref{eq: voltage phasor} and~\eqref{eq: current phasor}, we can see that each entry of the complex measurement vector $\zbf$ is linearly related to the complex system state vector $\xbf$. Therefore, the linear measurement equation, incorporating the measurement noise, can be obtained as follows:
\begin{equation}
\label{eq: PMU_measurements}
     \zbf = \Hbf\xbf + \ebf,
\end{equation}
 where $\Hbf\in\mathbb{C}^{m\times N}$ is a linear operator determined based on the network topology and line parameters according to~\eqref{eq: voltage phasor} and~\eqref{eq: current phasor}, and $\ebf\in\mathbb{C}^m$ is complex Gaussian noise. We do not require the assumption that $\Hbf$ is a full column rank matrix. In other words, we do not require state observability based on PMU measurements. 

\subsection{Attack model}
A GPS spoofing attack on a PMU can shift the time reference of the PMU, which the PMU uses to compute the phase angle measurements. Assuming that the frequencies of voltage and current waveforms are synchronized to the nominal frequency (e.g., 60 Hz in the United States), the bias in the time reference injected by the spoofing attack would cause a \emph{common} phase angle bias to all the phase angle measurements collected by the PMU\cite{jiang2013spoofing, fan2017synchrophasor}. 

Suppose that spoofing attack introduces a phase angle bias $\alpha_k$ to all of the phase angle measurements from PMU $k$ installed at bus $i$. Then, we can model the spoofed voltage and current measurements, denoted by $\bar{z}_{V_i}$ and $\bar{z}_{I_{il}}$ respectively, as follows:
\begin{subequations}
\label{eq: spoofed voltage phasor}
\begin{align}
\bar{z}_{V_i} & = e^{j\alpha_k}z_{V_i}\\
\bar{z}_{I_{il}} & = e^{j\alpha_k}z_{I_{il}},~l\in\Mmsc_i 
\end{align}
\end{subequations}
If we use $\zbf_k \in C^{m_k}$ and $\bar{\zbf}_k \in C^{m_k}$ to denote the intact measurements and the spoofed measurements from the PMU $k$ respectively, the spoofed measurements from PMU $k$ can be simply written as follows: 
\begin{equation}
    \bar{\zbf}_k = e^{j\alpha_k}\Ibf_{m_k}\zbf_k,
\end{equation}
where $\Ibf_{m_k}$ denotes the identity matrix of size $m_k\times m_k$.
This can be generalized to model PMU measurements from the entire PMU network as shown below,   
\begin{equation*}
\resizebox{0.9\hsize}{!}{$
\begin{bmatrix}
\bar{\zbf}_{1} \\
\bar{\zbf}_{2} \\
\vdots\\
\bar{\zbf}_{K} \\
\end{bmatrix}
= 
\begin{bmatrix}  
  e^{j\alpha_1}\Ibf_{m_1} & \bf{0} & \dots & \bf{0} \\
    \bf{0} & e^{j\alpha_2}\Ibf_{m_2} & \dots & \bf{0} \\
    \vdots & \vdots & \ddots & \vdots\\
    \bf{0} & \bf{0} & \dots & e^{j\alpha_K}\Ibf_{m_K}
  \end{bmatrix} \begin{bmatrix}
{\zbf}_1 \\
{\zbf}_2 \\
\vdots\\
{\zbf}_K \\
\end{bmatrix}$},
\end{equation*}
or equivalently, 
\begin{equation}
\begin{array}{ll}
\label{eq: attack_model2}
\bar{\zbf} & = \boldsymbol{\Phi}(\boldsymbol{\alpha})\zbf,
\end{array}
\end{equation}
where the diagonal matrix $\boldsymbol{\Phi}(\boldsymbol{\alpha})$ denotes the attack structure using $\boldsymbol{\alpha} \triangleq [\alpha_1, \alpha_2, \hdots, \alpha_K]^T$. The angle bias $\alpha_k$ is nonzero if the measurements from PMU $k$ are spoofed, and it is zero otherwise. This model is equivalent to the model employed in \cite{risbud2018vulnerability}. 

By combining~\eqref{eq: PMU_measurements} with~\eqref{eq: attack_model2} we can obtain the model for PMU measurements potentially subject to a GPS spoofing attack:
\begin{equation}
\label{eq: combined model}
\bar{\zbf} =\boldsymbol{\Phi}(\boldsymbol{\alpha})(\Hbf\xbf +\ebf)
\end{equation}
\subsection{Problem statement}
\label{sec:II_C}
Given a potentially spoofed measurement vector $\bar{\zbf}$ from the measurement model~\eqref{eq: combined model}, we aim to estimate $\boldsymbol{\alpha}$ such that we can recover the original measurement vector $\zbf$. Unfortunately, many $\boldsymbol{\alpha}$'s are fundamentally not identifiable from $\bar{\zbf}$. In particular, given noiseless PMU measurements $\bar{\zbf}$ generated from some attack $\boldsymbol{\alpha}$ and state $\xbf$, there can exist some $\bar{\boldsymbol{\alpha}}\neq \boldsymbol{\alpha}$ and $\bar{\xbf}$ satisfying:
\begin{equation}
 \bar{\zbf} =\boldsymbol{\Phi}(\bar{\boldsymbol{\alpha}})\Hbf\bar{\xbf},~~\textrm{or equivalently,  } \boldsymbol{\Phi}^{-1}(\bar{\boldsymbol{\alpha}})\bar{\zbf} \in \Rmsc({\Hbf}),
\end{equation}  
i.e., the measurements might be consistent with another attack scenario described by $\bar{\boldsymbol{\alpha}}$. Relying solely on the spoofed measurements $\bar{\zbf}$, it is impossible to detect which one is the true attack among the consistent attack scenarios. 

In order to alleviate this challenge and accommodate effective attack identification, we employ a practical assumption that only a few PMUs in $\Tmsc$ are spoofed by the adversary. In other words, $\boldsymbol{\alpha}$ is a sparse vector, or equivalently, only a few $\boldsymbol{\alpha}_k$'s are nonzero. By leveraging the sparse attack assumption, we aim to obtain an accurate estimate of $\boldsymbol{\alpha}$ based on observation of a potentially spoofed measurement vector $\bar{\zbf}$, which follows the measurement model~\eqref{eq: combined model}. Note that the state vector $\xbf$ is unknown. Once we obtain an estimate of $\boldsymbol{\alpha}$, we can use it in conjunction with the attack model~\eqref{eq: attack_model2} to recover the original measurement vector $\zbf$. 


Note that if the attack vector  $\boldsymbol{\alpha}$ is not sparse enough and its entries are designed in an elaborate manner, then  $\boldsymbol{\alpha}$ might not be fundamentally identifiable in the sparse error correction regime.
We formalize the attack identifiability in Section~\ref{sec: identifiability} and present a rigorous identifiability analysis of spoofing attacks. The results explain how the network topology and PMU locations affect fundamental identifiability of GPS spoofing attacks and therein can be leveraged to assess the vulnerability of power grid to GPS spoofing attacks by simply analyzing the grid topology and PMU locations (e.g., what is the minimum number of PMUs an attacker needs to spoof to be able to launch an unidentifiable attack). In Section~\ref{sec: algorithm}, we present a sparse error correction algorithm that can be used to effectively estimate spoofing attacks that are identifiable.

\section{Identifiability of sparse spoofing attacks}
\label{sec: identifiability}

Suppose there exist $\boldsymbol{\alpha}$ and $\bar{\boldsymbol{\alpha}}$ such that they are consistent with the noiseless PMU measurements, and $\bar{\boldsymbol{\alpha}}$ has a fewer number of nonzero entries than $\boldsymbol{\alpha}$. Since the sparse recovery algorithms inherently pick the most sparse solution to a problem~\cite{eldar2012compressed}, identifying the true sparse attack vector $\boldsymbol{\alpha}$ becomes fundamentally impossible in such a situation. Based on this intuition, the attack identifiability can be defined as below:
\begin{definition}
\label{def : global identifiability}
An attack $\boldsymbol{\alpha}$ is said to be \emph{identifiable} for a state $\xbf$ if there does \emph{not} exist $\bar{\boldsymbol{\alpha}}\neq \boldsymbol{\alpha}$ such that
\begin{enumerate}[(i)]
    \item $\|\bar{\boldsymbol{\alpha}}\|_0 \leq \|\boldsymbol{\alpha}\|_0 $, and
    \item $\boldsymbol{\Phi}(\boldsymbol{\alpha})\Hbf\xbf = \boldsymbol{\Phi}(\bar{\boldsymbol{\alpha}})\Hbf\bar{\xbf},$ for some $\bar{\xbf}$,~or equivalently,

    $\boldsymbol{\Phi}^{-1}(\bar{\boldsymbol{\alpha}})(\boldsymbol{\Phi}(\boldsymbol{\alpha})\Hbf\xbf) \in \Rmsc(\Hbf)
$
\end{enumerate}
\end{definition} 

Having formally defined identifiability, we perform the attack identifiability analysis to characterize the conditions for attack identifiability in terms of network topology, PMU locations and spoofed PMU locations. A major challenge in the identifiability analysis is that the spoofed measurement model~\eqref{eq: combined model} is nonlinear and involves complex-valued variables. To circumvent this challenge, we first introduce an alternative measurement vector $\bar{\wbf}$ that can be obtained by applying a transformation $T(\cdot)$ to $\bar{\zbf}$ and is linearly related to the voltage phase angles and the attack vector $\boldsymbol{\alpha}$. In the following proposition, we define the transformation $T(\cdot)$.

\begin{proposition}
\label{prop : alternative measurements mapping}
Let $\bar{\wbf}$ be a real-valued vector consisting of the following quantities:
\begin{center}
$\{\angle \bar{z}_{V_i}, \Delta \theta_{il}, \textrm{ for all } i\in \Tmsc \textrm{ and } l\in \Mmsc_i\}$,
\end{center}
where $\Delta\theta_{il} = (\theta_i - \theta_l)$ and $\theta_i$ denotes the voltage state angle at bus $i$. Then, there exists a mapping $T(.)$ such that $\bar{\wbf} = T(\bar{\zbf})$. Specifically, the entries of $\bar{\wbf}$ can be obtained from $\bar{\zbf}$ as follows:
\begin{subequations}
\label{eq: alternative measurements}
\begin{align}
\bar{w}_{\angle V_i} & = \angle \bar{z}_{V_i}\\
\bar{w}_{\Delta \theta_{il}} & = \angle (\frac{ (y_{il}^* - j\frac{b^s_{il}}{2})|\bar{z}_{V_i}|^2 - \bar{z}_{V_i}\bar{z}_{I_{il}}^*}{y_{il}^*}) 
\end{align}
\end{subequations}
\end{proposition}
\begin{proof}
See Appendix~\ref{app:alt_measurement}.
\end{proof}

Unlike the original measurement vector $\bar{\zbf}$, the alternative measurement vector $\bar{\wbf} = T(\bar{\zbf})$ can be shown to be linearly related to the voltage phase angle vector $\boldsymbol{\theta}\in \mbbR^{N}$ which comprises all $\theta_i,~\forall i\in \Vmsc$, and the attack vector $\boldsymbol{\alpha}$. First, we can use~\eqref{eq: voltage phasor} and~(\ref{eq: spoofed voltage phasor}a) to derive that, 
\begin{equation}  
\label{eq: spoofed voltage angle}
\begin{aligned}
\bar{w}_{\angle V_{i}} & = \angle \bar{z}_{V_i} = \angle (e^{j\alpha_k}z_{V_i}) = \angle |x_i|e^{j(\theta_i+\alpha_k)} = \alpha_k + \theta_i,
\end{aligned}
\end{equation}
where $\boldsymbol{\alpha}_{k}$ is the phase angle bias introduced by the attack on PMU $k$ installed in bus $i$. By concatenating $\bar{w}_{\angle V_i}$ for all the buses with PMUs, we obtain the vector $\bar{\wbf}_{\angle V}$, which is the vector of voltage phase angle measurements in the presence of a spoofing attack:
\begin{equation}
    \bar{\wbf}_{\angle V} =\Hbf_{\angle V}\boldsymbol{\theta} + \boldsymbol{\alpha}. 
    \label{eq: wv model}
\end{equation}
The matrix $\Hbf_{\angle V}\in\mbbR^{K\times N}$ is determined using~(\ref{eq: spoofed voltage angle}), where each row corresponds to a voltage angle measurement, and each column corresponds to a particular bus in the network. Suppose that row $j$ of $\Hbf_{\angle V}$ corresponds to the measurement $\bar{w}_{\angle V_i}$ from the PMU in bus $i$. Then entry $(j,i)$ in $\Hbf_{\angle V}$ is set to one and the rest of the entries in row $j$ are set to zeros.

Similarly, from the definition of $\bar{w}_{\Delta \theta_{il}}$ in Proposition~\ref{prop : alternative measurements mapping},
\begin{equation}
\label{eq: alternative angle difference}
    \bar{w}_{\Delta \theta_{il}} = \theta_i - \theta_l
\end{equation}
Thus, concatenating $\bar{w}_{\Delta_{\theta{il}}}$ for all $i\in\Tmsc$ and $l\in\Mmsc_i$, we obtain the vector $\bar{\wbf}_{\Delta}$ consisting of voltage angle differences across all the lines measured by PMUs, which can be mathematically written as below:
\begin{equation}
        \bar{\wbf}_{\Delta} =\Hbf_{\Delta}\boldsymbol{\theta}
        \label{eq: w_delta}
\end{equation}
Each row of $\Hbf_{\Delta}\in\mbbR^{(m-K)\times N}$ corresponds to voltage angle difference measurements computed using~\eqref{eq: alternative angle difference} and each column corresponds to a particular bus in the network. Suppose that row $j$ of $\Hbf_{\Delta}$ corresponds to $\bar{w}_{\Delta \theta_{il}}$ for some bus $i\in\Tmsc$ and $l\in\Mmsc_i$. Then all the entries of row $j$ are set to zeros except for $(j,i)$ and $(j,l)$ entries of the matrix $\Hbf_{\Delta}$, which are set to one and negative one, respectively. 

Hence now we can write the alternative measurement model using~\eqref{eq: wv model} and~\eqref{eq: w_delta} as follows.
\begin{equation}
\label{eq: wbar model}
    T(\bar{\zbf}) = \bar{\wbf} = \begin{bmatrix}
    \bar{\wbf}_{\angle V}\\
    \bar{\wbf}_{\Delta}
    \end{bmatrix}
    =
    \begin{bmatrix}
    \Hbf_{\angle V}\\
    \Hbf_{\Delta}
    \end{bmatrix}
    \boldsymbol{\theta}
     + 
     \begin{bmatrix}
     \boldsymbol{\alpha}\\
     \boldsymbol{0}
     \end{bmatrix}
\end{equation}

\subsection{Identifiability analysis}
In this section, we perform identifiability analysis based on the alternative measurement model derived above. Recall that identifiability is defined by Definition~\ref{def : global identifiability}, which states that if a GPS spoofing attack $\boldsymbol{\alpha}$ is \emph{not} identifiable from measurements $\bar{\zbf}$ generated by $(\boldsymbol{\alpha}, \xbf)$, then there exists $\bar{\boldsymbol{\alpha}}\neq \boldsymbol{\alpha}$ such that $\|\bar{\boldsymbol{\alpha}}\|_0 \leq \|\boldsymbol{\alpha}\|_0 $ and, 
\begin{equation}
\label{eq: prior to transformation}
\bar{\zbf} = \boldsymbol{\Phi}(\boldsymbol{\alpha})\Hbf\xbf = \boldsymbol{\Phi}(\bar{\boldsymbol{\alpha}})\Hbf\bar{\xbf},    
\end{equation}
for some $\bar{\xbf}$. Hence, by applying transformation $T(.)$ defined in Proposition~\ref{prop : alternative measurements mapping} on~\eqref{eq: prior to transformation}, we can see that,
\begin{center}
$T(\boldsymbol{\Phi}(\boldsymbol{\alpha})\Hbf\xbf) = T(\boldsymbol{\Phi}(\bar{\boldsymbol{\alpha}})\Hbf\bar{\xbf}),$
\end{center}
or equivalently,
\begin{equation}
\label{eq: wbar}
\begin{aligned}
 \begin{bmatrix}
    \Hbf_{\angle V}   \\
    \Hbf_{\Delta} 
    \end{bmatrix}
      \boldsymbol{\theta}
      + 
      \begin{bmatrix}
          \boldsymbol{\alpha}  \\
    \boldsymbol{0}
    \end{bmatrix}
=
\begin{bmatrix}
    \Hbf_{\angle V}   \\
    \Hbf_{\Delta} 
    \end{bmatrix}
      \bar{\boldsymbol{\theta}}
      + 
      \begin{bmatrix}
          \bar{\boldsymbol{\alpha}}  \\
    \boldsymbol{0}
    \end{bmatrix},
\end{aligned}
\end{equation}
where $\bar{\boldsymbol{\theta}}$ denotes the angles of the state $\bar{\xbf}$, precisely $\bar{\theta}_i = \angle{\bar{x}_i}$. 
This implies that if the attack $\boldsymbol{\alpha}$ is not identifiable, then there exists $\bar{\boldsymbol{\alpha}}\neq \boldsymbol{\alpha}$ such that $\|\bar{\boldsymbol{\alpha}}\|_0 \leq \|\boldsymbol{\alpha}\|_0 $ and 
\begin{equation}
\label{eq: identifiability def alternative model}
\begin{aligned}
 \begin{bmatrix}
    \Hbf_{\angle V}   \\
    \Hbf_{\Delta} 
    \end{bmatrix}
      \boldsymbol{(\bar{\theta} - \theta)}
      =
      \begin{bmatrix}
          \boldsymbol{\alpha} - \bar{\boldsymbol{\alpha}}  \\
    \boldsymbol{0}
    \end{bmatrix}
\end{aligned},~\textrm{for some }\boldsymbol{\theta} \textrm{ and }\boldsymbol{\bar{\theta}}.
\end{equation}

Since this implies that $(\boldsymbol{\bar{\theta} - \theta})$ is in the null space of $\Hbf_\Delta$,~\eqref{eq: identifiability def alternative model} is equivalent to, 
\begin{equation}
\label{eq: hvbdelta and alpha}
          \boldsymbol{\alpha} - \bar{\boldsymbol{\alpha}} 
           \in \Rmsc(\Hbf_{\angle V}\Bbf_{\Delta}),
\end{equation}
where $\Bbf_\Delta$ forms a basis for $\Nmsc(\Hbf_\Delta)$. The contrapositive of this statement directly induces the following proposition.
\begin{proposition}
\label{prop : definition global identifiability}
An attack $\boldsymbol{\alpha}$ is identifiable for any state $\xbf$ if there does not exist $\bar{\boldsymbol{\alpha}}\neq \boldsymbol{\alpha}$ such that
\begin{enumerate}[(i)]
    \item  $\|\bar{\boldsymbol{\alpha}}\|_0 \leq \|\boldsymbol{\alpha}\|_0 $
    \vspace{0.5em}
    \item $ 
        \boldsymbol{\alpha} -  \bar{\boldsymbol{\alpha}}
       \in \Rmsc(\Hbf_{\angle V}\Bbf_{\Delta})$
\end{enumerate}
\end{proposition}
This proposition provides a sufficient condition for identifiability of $\boldsymbol{\alpha}$ in terms of the column space of $\Hbf_{\angle V}\Bbf_{\Delta}$. Candes and Tao presented in~\cite{candes2005decoding} a theoretical result that can be used to further simplify the sufficient condition in Proposition~\ref{prop : definition global identifiability} to a condition in terms of the sparsity of $\boldsymbol{\alpha}$. By applying Lemma 1.7 in~\cite{candes2005decoding} to the conditions in Proposition~\ref{prop : definition global identifiability}, we can obtain the following lemma.
\begin{lemma}
\label{lemma : global cospark}
An attack $\boldsymbol{\alpha}$ is identifiable for any state $\xbf$ if,
\begin{center}
$\|\boldsymbol{\alpha}\|_0  < \frac{1}{2}$Cospark$(\Hbf_{\angle V}\Bbf_{\Delta}) $,
\end{center}
where cospark of a matrix $\Abf$ is defined as,
\begin{equation*}
   \textrm{Cospark}(\Abf) = \min\limits_{\hbf\in\Rmsc{(\Abf)},\hbf\neq\boldsymbol{0}} \|\hbf\|_0 
\end{equation*}
\end{lemma}
\begin{proof}
 See Appendix~\ref{proof: lemma : global cospark}.
\end{proof}
Even though finding the cospark of a matrix is generally an NP-hard problem, the special structures of the matrices $\Hbf_{\Delta}$ and $\Hbf_{\angle V}$ make it possible to derive the cospark of $\Hbf_{\angle V}\Bbf_{\Delta}$ exactly in terms of the locations and the number of PMUs in the network. In order to understand this underlying structure of the matrices, we first define the concept of a \emph{zone} in the power network, which we identify as a region of the network measured by a subset of PMUs whose measurements are correlated via sharing of some common latent state variables.

\tikzset{main node/.style={circle,draw,minimum size=0.38cm,inner sep=0pt},}
\begin{figure}
\centering
\subfloat[$\Gmsc$]{\begin{tikzpicture}
\label{fig:5bus network-a}
    \node[main node] (1) {$1$};
    \node[main node] (2) [below left = 0.5cm and 1 cm of 1, fill = black!30]  {$2$};
    \node[main node] (3) [below right = 0.8cm and 0.8cm of 1] {$3$};
    \node[main node] (4) [fill = black!30, right = 0.8cm of 1, fill = black!30]{$4$};
    \node[main node] (5) [fill = black!30, above right = 0.15cm and 0.75cm of 3, fill = black!30]{$5$};
    \path[draw,thick]
    (1) edge node {} (2)
    (1) edge node {} (3)
    (3) edge node {} (5)
    (1) edge node {} (4); 
    \end{tikzpicture}}\hfill
\subfloat[$\Gmsc_\Tmsc$]{\begin{tikzpicture}
\label{fig:5bus network-b}
    \node[main node] (1) [draw = blue!30, thick]{$1$};
    \node[main node] (2) [draw = blue!30, thick, below left = 0.5cm and 1 cm of 1, fill = blue!30]  {$2$};
    \node[main node] (3) [draw = red!30, thick,below right = 0.8cm and 0.8cm of 1] {$3$};
    \node[main node] (4) [draw = blue!30, thick,right = 0.8cm of 1, fill = blue!30]{$4$};
    \node[main node] (5) [draw = red!30, thick,above right = 0.15cm and 0.75cm of 3, fill = red!30]{$5$};
    \path[draw,thick]
    (1) edge[blue!30] node {} (2)
    (3) edge[red!30] node {} (5)
    (1) edge[blue!30] node {} (4); 
    \end{tikzpicture}}\quad
    \resizebox{0.49\hsize}{!}{
$\Hbf_{\Delta} =
\begin{array}{c c c;{2pt/2pt} c c c }
\kbordermatrix{
 & (1) & (2) & (4)~\vdots & (3) & (5)\\
(\bar{w}_{\Delta \theta_{21}}) &-1 & 1 & 0~~\vdots & 0 & 0\\
(\bar{w}_{\Delta \theta_{41}}) &-1 & 0 & 1~~\vdots & 0 & 0 \\ \hdashline[2pt/2pt]
(\bar{w}_{\Delta \theta_{53}})   & 0 & 0 & 0~~\vdots & -1 & 1\\ }
\end{array}$}
\resizebox{0.49\hsize}{!}{
$\Hbf_{\angle V} =
\begin{array}{c c c;{2pt/2pt} c c c }
\kbordermatrix{
 & (1) & (2) & (4)~\vdots & (3) & (5)\\
(\bar{w}_{\angle V_2}) &0 & 1 & 0~~\vdots & 0 & 0\\
(\bar{w}_{\angle V_4}) &0 & 0 & 1~~\vdots & 0 & 0 \\ \hdashline[2pt/2pt]
(\bar{w}_{\angle V_5})   & 0 & 0 & 0~~\vdots & 0 & 1\\ 
     }
\end{array}$}
\resizebox{0.49\hsize}{!}{
$\Bbf_{\Delta}^T = 
\begin{array}{c c c;{2pt/2pt} c c c }
\kbordermatrix{
 & (1) & (2) & (4) & (3) & (5)\\
(\textrm{zone }1) &    1 & 1 & 1 & 0 & 0 \\
(\textrm{zone }2) &    0 & 0 & 0 & 1 & 1  \\     
}
\end{array}$}
    \caption{Illustration of zones using a simple power network. The PMUs deployed on $\Tmsc = \{2, 4, 5\}$ measure outgoing current flow from 2 to 1, 4 to 1 and 5 to 3. Measurement graph $\Gmsc_\Tmsc = (\{1,2,3,4,5\},\{(2,1),(4,1),(3,5)\})$, consists of two zones where $\Vmsc_{\Tmsc}^{(1)} = \{1, 2, 4\}$ and $\Vmsc_{\Tmsc}^{(2)} = \{3, 5\}$.}
\label{fig_sim}
\end{figure}
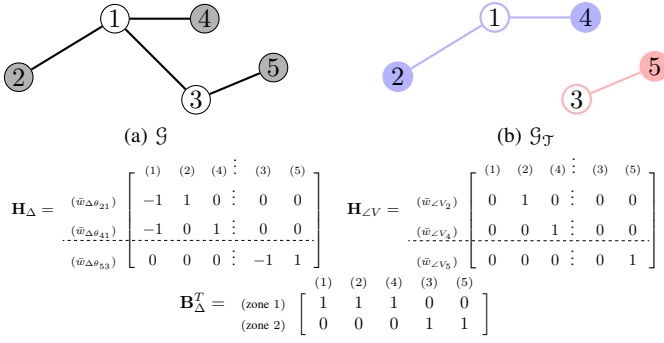

\textbf{Zones in a Power network:}
For each $i\in\Tmsc$, we define the graph $\Gmsc_i = (\Vmsc_i, \Emsc_i)$ such that $\Vmsc_i = \{i\}\cup\Mmsc_i$ and $\Emsc_i\subset\Emsc$ consists of the edges $\{i,l\},~\forall~l\in \Mmsc_i$. In other words, $\Gmsc_i$ is the subgraph of the topology consisting of all the lines measured by the PMU located at bus $i$. Then, we use $\Gmsc_\Tmsc$ to denote the union of all $\Gmsc_i$'s with $i\in\Tmsc$, \ie, $\Gmsc_\Tmsc  = (\Vmsc_\Tmsc, \Emsc_\Tmsc) = (\underset{j\in\Tmsc}{\cup}\Vmsc_j, \underset{j\in\Tmsc}{\cup}\Emsc_j)$. Since $\Gmsc_\Tmsc$ depicts the region in the power network measured by the PMUs placed in the set of buses $\Tmsc$, we refer $\Gmsc_\Tmsc$ as the measurement graph of PMUs in $\Tmsc$. Figure~\ref{fig:5bus network-a} and Figure~\ref{fig:5bus network-b} illustrate a simple power network with five buses, four branches, and three PMUs, and the measurement graph $\Gmsc_\Tmsc$ where $\Tmsc=\{2,4,5\}$, respectively.

Given the measurement graph $\Gmsc_\Tmsc$, we define \emph{zones} as connected components of graph $\Gmsc_\Tmsc$ as follows. We say that bus $i$ and bus $j$ in $\Gmsc_\Tmsc$ are \emph{reachable} from each other if there exists a path\footnote{Path is a sequence of distinct edges that join a sequence of distinct vertices.} in $\Gmsc_\Tmsc$ that has $i$ and $j$ as the end points. We partition the vertex set $\Vmsc_\Tmsc$ into $\Vmsc_{\Tmsc}^{(1)},\dots,\Vmsc_{\Tmsc}^{(\Gamma)}$ such that two vertices $i$ and $j$ are in the same $\Vmsc_{\Tmsc}^{(\gamma)}$ if and only if they are reachable from each other. We refer to the set of buses in $\Vmsc_{\Tmsc}^{(\gamma)}$ as \emph{Zone $\gamma$}. For instance, in Figure~\ref{fig:5bus network-b}, we have two zones $\Vmsc_{\Tmsc}^{(1)}$ and $\Vmsc_{\Tmsc}^{(2)}$; each of them corresponds to one of the two connected components of $\Gmsc_\Tmsc$. This partitioning in the vertex domain naturally partition the set of PMUs $\Tmsc$ into $\Tmsc^{(1)},\dots,\Tmsc^{(\Gamma)}$, where we use $\Tmsc^{(\gamma)}$ to denote the set of buses with PMUs in Zone $\gamma$,~\ie, $\Tmsc^{(\gamma)} = \Tmsc \cap\Vmsc_{\Tmsc}^{(\gamma)}$. Note that the concept of zone defined in this paper is different from the more popular concept of observable island~\footnote{An observable island is a region in the network that makes all the states in the region observable from the measurements collected within the region}\cite{baldwin1993power,monticelli1985network}; specifically, a single observable island can contain multiple zones. For instance, the entire network in Figure~\ref{fig:5bus network-a} is observable based on PMU measurements, but as Figure~\ref{fig:5bus network-b} shows this observable island contains two zones.

From the PMU measurement model defined in~\eqref{eq: voltage phasor} and~\eqref{eq: current phasor}, the PMU measurements from Zone $\gamma$ denoted by $\bar{\zbf}^{(\gamma)}$, depend only on state variables associated with buses in Zone $\gamma$, which we denote by $\xbf^{(\gamma)}$. Therefore, the spoofed PMU measurement model~\eqref{eq: attack_model2} and~\eqref{eq: combined model} has the following block structure:
\begin{equation}
\label{eq: decoupled model}
\resizebox{1\hsize}{!}{
$\begin{bmatrix}
 \bar{\zbf}^{(1)} \\
 \vdots\\
 \bar{\zbf}^{(\Gamma)} 
\end{bmatrix}
 =\begin{bmatrix}
 \boldsymbol{\Phi}_1(\boldsymbol{\alpha}^{(1)}) & \dots & \obf\\
 \vdots & \ddots &\vdots\\
 \obf & \dots & \boldsymbol{\Phi}_\Gamma(\boldsymbol{\alpha}^{(\Gamma)}) 
 \end{bmatrix}
 \begin{bmatrix}
  \Hbf^{(1)} & \dots & \obf\\
 \vdots & \ddots &\vdots\\
 \obf & \dots &  \Hbf^{(\Gamma)}
 \end{bmatrix}
 \begin{bmatrix}
 \xbf^{(1)} \\
 \vdots\\
 \xbf^{(\Gamma)} 
\end{bmatrix}$},
\end{equation}  
where $\boldsymbol{\alpha}^{(\gamma)}$ denotes the sub-vector of $\boldsymbol{\alpha}$ that represents the angle biases introduced to the PMUs in Zone $\gamma$. Moreover, $\boldsymbol{\Phi}_\gamma(\boldsymbol{\alpha}^{(\gamma)})$ is a diagonal matrix which denotes the submatrix of $\boldsymbol{\Phi}(\boldsymbol{\alpha})$ corresponding to Zone $\gamma$, and $\Hbf^{(\gamma)}$ denotes the submatrix of $\Hbf$ that represents the linear relation between PMU measurements and states in Zone $\gamma$. The decomposed spoofed measurement model per zone can be given as below:
\begin{equation}
\label{eq: decompose model zonewise}
 \bar{\zbf}^{(\gamma)} =\boldsymbol{\Phi}_\gamma(\boldsymbol{\alpha}^{(\gamma)})\Hbf^{(\gamma)}\xbf^{(\gamma)}.
\end{equation}  
Similarly, the linear model~\eqref{eq: wbar model} of the alternative measurements can be decomposed in to zones as follows:
\begin{equation}
\label{eq: decoupled alternative model2}
\resizebox{0.75\hsize}{!}{$
\begin{bmatrix}
 \bar{\wbf}^{(1)}_{\angle V} \\
  \vdots\\
  \bar{\wbf}^{(\Gamma)}_{\angle V}\\
 \bar{\wbf}^{(1)}_{\Delta} \\
 \vdots\\
 \bar{\wbf}^{(\Gamma)}_{\Delta}\\ 
\end{bmatrix}
 = \begin{bmatrix}
  \Hbf^{(1)}_{\angle V} & \dots & \obf\\
 \vdots & \ddots &\vdots\\
 \obf & \dots &  \Hbf^{(\Gamma)}_{\angle V}\\
  \Hbf^{(1)}_{\Delta} & \dots & \obf\\
 \vdots & \ddots &\vdots\\
 \obf & \dots &  \Hbf^{(\Gamma)}_{\Delta}\\
 \end{bmatrix}
 \begin{bmatrix}
 \boldsymbol{\theta}^{(1)} \\
 \vdots\\
 \boldsymbol{\theta}^{(\Gamma)} 
\end{bmatrix},
+
\begin{bmatrix}
 \boldsymbol{\alpha}^{(1)} \\
 \vdots\\
 \boldsymbol{\alpha}^{(\Gamma)} \\
 \boldsymbol{0} \\
 \vdots\\
 \boldsymbol{0} \\
\end{bmatrix},$}
\end{equation}  
where $\Hbf^{(\gamma)}_{\angle V}\in\mbbR^{K^{(\gamma)}\times N^{(\gamma)}}$ and  $\Hbf^{(\gamma)}_{\Delta} \in \mbbR^{({m^{(\gamma)}-K^{(\gamma)}})\times N^{(\gamma)}}$ denote the submatrices of $\Hbf_{\angle V}$ and $\Hbf_{\Delta}$ that represent the linear relation of alternative measurements $\bar{\wbf}_{\angle V}^{(\gamma)}$ and $\bar{\wbf}_{\Delta}^{(\gamma)}$ with voltage state angles in Zone $\gamma$, respectively. Note that $m^{(\gamma)}, K^{(\gamma)}$ and $N^{(\gamma)}$ represents the number of measurements in Zone $\gamma$, the number of PMUs in Zone $\gamma$ and the number of busses in Zone $\gamma$ respectively. The decomposed alternative measurement model~\eqref{eq: decoupled alternative model2} implies that $\Hbf_{\angle V}$ and $\Hbf_{\Delta}$ are block matrices with  off-diagonal blocks equal to zero and each diagonal block corresponds to a zone in the network.

Due to this special block structure of $\Hbf_\Delta$ described in~\eqref{eq: decoupled alternative model2} and the sparsity pattern of each block in this matrix imposed by~\eqref{eq: w_delta}, we can derive the basis for the null space of $\Hbf_\Delta$ as given in the following proposition:   
\begin{proposition}
\label{prop: bdelta}
The basis for $\Nmsc(\Hbf_\Delta)$ is a block matrix $\Bbf_\Delta\in\mbbR^{N\times \Gamma}$, 
\begin{equation*}
\resizebox{0.5\hsize}{!}{$
     \Bbf_{\Delta} =
    \begin{bmatrix}
    \Bbf_{\Delta}^{(1)} & \boldsymbol{0} & \hdots \boldsymbol{0} \\
    \boldsymbol{0} & \Bbf_{\Delta}^{(2)} & \hdots \boldsymbol{0}  \\
    \vdots & \hdots&  \vdots\\
    \boldsymbol{0} & \boldsymbol{0} & \hdots \Bbf_{\Delta}^{(\Gamma)}  
    \end{bmatrix}$},
    \textrm{ where }     \Bbf_\Delta^{(\gamma)} = \boldsymbol{1}_{N^{(\gamma)}}.
\end{equation*}
Here $\boldsymbol{1}_{N^{(\gamma)}}$ denotes the $N^{(\gamma)}-$ dimensional vector with all entries equal to one.
\end{proposition}
\begin{proof}
See Appendix~\ref{proof: prop: bdelta}.
\end{proof}
Since both $\boldsymbol{\Hbf}_{\angle V}$ and $\boldsymbol{\Bbf}_{\Delta}$ are matrices with the special block structure, $\Hbf_{\angle V} \Bbf_\Delta$ also takes the same block structure,~\ie,
\begin{equation}
\resizebox{0.8\hsize}{!}{$
\label{eq: structure of H v B delta }
     \Hbf_{\angle V}\Bbf_{\Delta} =
    \begin{bmatrix}
         \Hbf_{\angle V}^{(1)}\Bbf_{\Delta}^{(1)} & \boldsymbol{0} & \hdots \boldsymbol{0} \\
    \boldsymbol{0} &      \Hbf_{\angle V}^{(2)}\Bbf_{\Delta}^{(2)} & \hdots  \boldsymbol{0}  \\
    \vdots & \hdots&  \vdots\\
    \boldsymbol{0} & \boldsymbol{0} & \hdots  \Hbf_{\angle V}^{(\Gamma)}\Bbf_{\Delta}^{(\Gamma)} 
    \end{bmatrix},$}
\end{equation}
where $\Hbf_{\angle V}^{(\gamma)} \Bbf_\Delta^{(\gamma)}$ denotes the diagonal block corresponding to Zone $\gamma$. By leveraging this special block structure of $\Hbf_{\angle V} \Bbf_\Delta$, we can derive the following lemma:
\begin{lemma}
\label{lemma : cospark in terms of Kmin}
\begin{center}
~~~~~~~~~~~~~~~~~~~~~~~~~~~~~~~~~~~~~~~~~~~~~~~~~~~~~\\
Cospark $(\Hbf_{\angle V}\Bbf_{\Delta})$ = $\min\limits_{\gamma\in \{1,2,\dots,\Gamma\}}\textrm{Cospark}(\Hbf_{\angle V}^{(\gamma)}\Bbf_{\Delta}^{(\gamma)})$
\end{center}
\end{lemma}
\begin{proof}
 See Appendix~\ref{proof: lemma : cospark in terms of Kmin}. 
\end{proof}
Furthermore, using Proposition~\ref{prop: bdelta} together with the structure of $\Hbf_{\angle V}$ imposed by~\eqref{eq: decoupled alternative model2}, we can obtain the cospark of $\Hbf_{\angle V}^{(\gamma)}\Bbf_{\Delta}^{(\gamma)}$ in terms of the number of PMUs in each zone.
\begin{lemma}
\label{lemma : cospark in terms of Kmin2} For all zones $\gamma \in \{1,2,\dots,\Gamma\}$ in the network,
\begin{center}
Cospark $(\Hbf_{\angle V}^{(\gamma)}\Bbf_{\Delta}^{(\gamma)}) $ = $K^{(\gamma)}$
\end{center}
\end{lemma}
\begin{proof}
 See Appendix~\ref{proof: lemma : cospark in terms of Kmin2}. 
\end{proof}

Combining Lemma~\ref{lemma : global cospark}, Lemma~\ref{lemma : cospark in terms of Kmin} and Lemma~\ref{lemma : cospark in terms of Kmin2} we obtain a sufficient condition for the identifiability of $\boldsymbol{\alpha}$ in terms of $K_{min}$, which is the smallest number of PMUs in any zone of the power network.
\begin{theorem}
\label{thm : condition on sparsity of alpha - global}
An attack $\boldsymbol{\alpha}$ is identifiable for any state $\xbf$ if, \begin{center}
$\|\boldsymbol{\alpha}\|_0 \leq \ceil{\frac{K_{min}}{2} - 1}  $,
\end{center}
where $K_{min} = \min\limits_{\gamma\in \{1,2,\dots,\Gamma\}}K^{(\gamma)}$.\\
In addition, given any state $\xbf$, there exists an unidentifiable attack $\boldsymbol{\alpha}$ for $\xbf$ with the sparsity level $\|\boldsymbol{\alpha}\|_0  = \ceil{\frac{K_{min}}{2} - 1} + 1$.
\end{theorem}
\begin{proof}
 See Appendix~\ref{proof: thm : condition on sparsity of alpha - global}.
\end{proof}
Suppose we measure the size of a zone in the network by the number of PMUs in it. Theorem~\ref{thm : condition on sparsity of alpha - global}  implies that if the number of spoofed PMUs in the entire network is less than half of the number of PMUs in the smallest zone, such attacks are identifiable. Furthermore, we have proved that if the number of spoofed PMUs exceed this threshold by 1, then there exists an unidentifiable attack. By leveraging the block structure of $\Hbf_{\angle V}\Bbf_\Delta$~\eqref{eq: structure of H v B delta } and decomposability of the model~\eqref{eq: decoupled alternative model2}, we were able to derive the following theorem providing us with a more relaxed condition that implies identifiability of the attack:
\begin{theorem}
\label{thm : condition on sparsity of alpha - local}
An attack $\boldsymbol{\alpha}$ is identifiable for any state $\xbf$ if, 
\begin{center}
$\|\boldsymbol{\alpha}^{(\gamma)}\|_0 \leq \ceil{\frac{K^{(\gamma)}}{2} - 1}~~\forall \gamma\in\{1,\dots,\Gamma\}$.
\end{center}
In addition, given any state $\xbf$ there exists an unidentifiable attack $\boldsymbol{\alpha}$ that has the sparsity level $\|\boldsymbol{\alpha}^{(\bar{\gamma})}\|_0  = \ceil{\frac{K^{(\bar{\gamma})}}{2} - 1} + 1$, for some $\bar{\gamma}\in\{1,2,\dots,\Gamma\}$ and $\|\boldsymbol{\alpha}^{(\gamma)}\|_0  \leq \ceil{\frac{K^{(\gamma)}}{2} - 1}$ for $\gamma\in\{1,2,\dots,\Gamma\}\setminus\{\bar{\gamma}\}$.
\end{theorem}
\begin{proof}
See Appendix~\ref{proof: thm : condition on sparsity of alpha - local}.
\end{proof}
This theorem states that as long as the number of spoofed PMUs in each zone is less than half of the number of PMUs in the zone, the attack is identifiable. Furthermore if there exists at least one zone where the sparsity condition is not satisfied then there exists an unidentifiable attack.

  \begin{figure}
     \centering
     \includegraphics[width=0.45\textwidth]{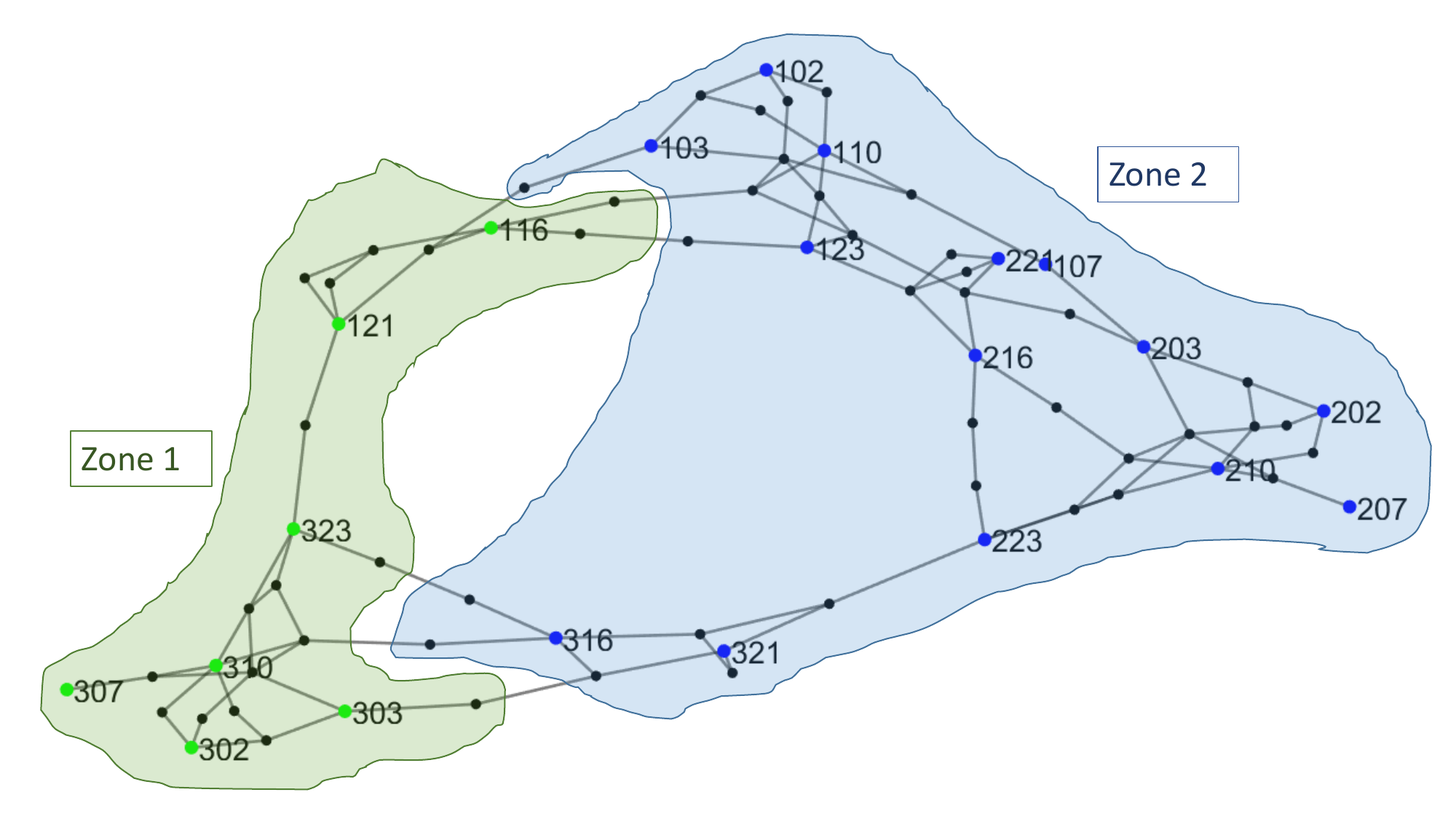}
     \caption{Observable PMU placement in IEEE RTS-96 test network and the two zones induced by these PMUs. PMUs in Zone 1 and Zone 2 are indicated by green and blue dots respectively, and their corresponding bus numbers.}
     \label{fig:RTS96}
 \end{figure}
 
\textbf{Leveraging identifiability analysis to improve grid resilience:}
Theorem~\ref{thm : condition on sparsity of alpha - global} implies that the smaller $K_{min}$, the more vulnerable the grid is to spoofing attacks in that the attacker can launch an unidentifiable spoofing attack by spoofing a smaller number of PMUs. This implies that when we allocate PMUs (or add an additional PMU to the grid), we can improve the grid resilience by maximizing $K_{min}$,~\ie, the number of PMUs in the zone containing the smallest number of PMUs. 

Figure~\ref{fig:RTS96} illustrates this idea with an example PMU allocation for the RTS-96 test network. This PMU allocation has 21 PMUs that naturally induces two zones in the network, with 7 and 14 PMUs respectively, where $K_{min} = 7$. Based on Theorem~\ref{thm : condition on sparsity of alpha - global}, if the number of spoofed PMUs is less than or equal to three, then the attack is identifiable regardless of the locations of the spoofed PMUs. Suppose that the operator combines Zone 2 with Zone 1 by deploying an additional PMU such that $K_{min}$ will be increased to 22. The new PMU placement ensures that the network is resilient to any spoofing attack with less than 11 spoofed PMUs. Thus the operators can significantly reduce the vulnerability of the network to spoofing attacks at a small increase in the cost.


\section{PMU data correction algorithm}
\label{sec: algorithm}

\begin{figure}[t]
 \centering
 \includegraphics[width=0.35\textwidth]{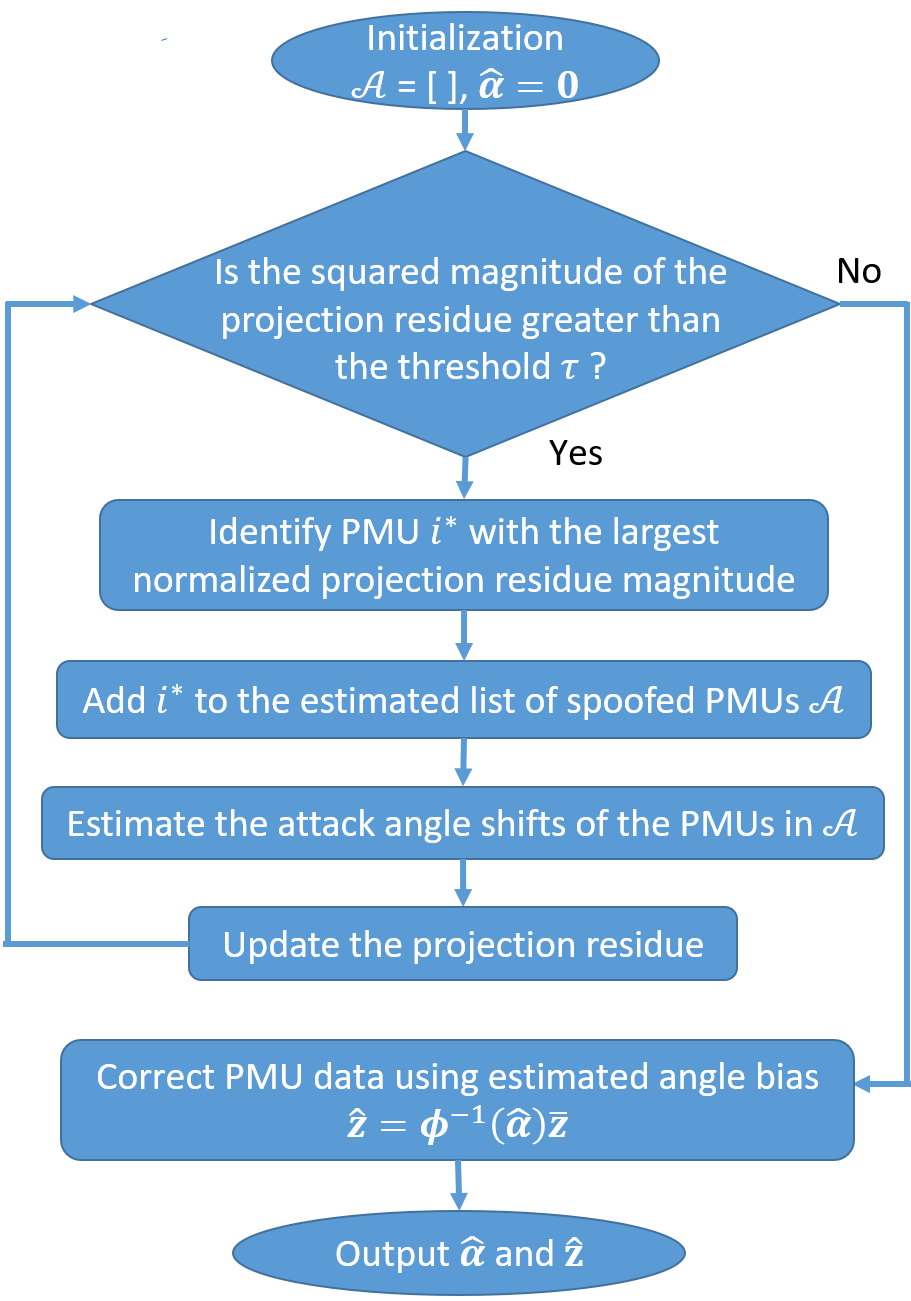} \caption{Flow chart of the sparse GPS spoofing correction algorithm}
 \label{fig: Flow chart}
\end{figure}

In this section, we present a sparse error correction algorithm to mitigate GPS spoofing. Note that we can rewrite the spoofed measurement model~\eqref{eq: combined model} as follows:
\begin{equation}
\label{eq: model}
\boldsymbol{\Phi}^{-1}(\boldsymbol{\alpha})\bar{\zbf} = \Hbf\xbf + \ebf.
\end{equation}
The above equation implies that with the true $\boldsymbol{\alpha}$, $\boldsymbol{\Phi}^{-1}(\boldsymbol{\alpha})\bar{\zbf}$ would reside very close to the column space of $\Hbf$, where the distance will be due to a small perturbation introduced by the measurement noise. In particular, if we project $\boldsymbol{\Phi}^{-1}(\boldsymbol{\alpha})\bar{\zbf}$  onto $\Rmsc(\Hbf)$, the projection residue $\rbf$ can be given as below, 
\begin{equation}
\label{eq: proj residue}
    \rbf := (\textbf{I}_{m} - \Pbf_{\Hbf})\boldsymbol{\Phi}^{-1}(\boldsymbol{\alpha})\bar{\zbf},
\end{equation}
where $\Pbf_{\Hbf}$ denotes the projection operator for projection on to the column space of $\Hbf$. Then, $\rbf$ is equivalent to the projection of only the measurement noise $\ebf$ onto the orthogonal complement of $\Rmsc(\Hbf)$ because, by plugging~\eqref{eq: model} into~\eqref{eq: proj residue},
\begin{equation}
\label{eq: proj residue2}
    \rbf = (\textbf{I}_{m} - \Pbf_{\Hbf})(\Hbf\xbf + \ebf) = (\textbf{I}_{m} - \Pbf_{\Hbf})\ebf~.
\end{equation}

We propose to estimate  $\boldsymbol{\alpha}$ by finding the sparsest estimate of  $\boldsymbol{\alpha}$ that makes the squared magnitude of the projection residue~\eqref{eq: proj residue} no greater than a pre-set threshold $\tau$: 
\begin{equation}
        \label{eq: optimization_problem}
\begin{aligned}
    \hat{\boldsymbol{\alpha}} = & \argmin_{\boldsymbol{\alpha}} \Vert\boldsymbol{\alpha}\Vert_0\\
    & \textrm{subject to~~~} \Vert(\textbf{I}_{m} - \Pbf_{\Hbf})\boldsymbol{\Phi}^{-1}(\boldsymbol{\alpha})\bar{\zbf})\Vert_2^2 \leq \tau~,
\end{aligned}
\end{equation}
The threshold $\tau$ is set such that the probability of the event $\|(\textbf{I}_{m} - \Pbf_{\Hbf})\ebf\|_2^2\geq\tau$ is equal to the target false alarm rate, according to the distribution of the noise vector $\ebf$~\footnote{If the noise distribution is not known, it can be estimated using intact historic PMU data.}.

Note that by solving the above optimization, we aim to localize the spoofed PMUs and estimate the phase angle biases of the spoofed measurements at the same time. The approach in~\cite{fan2017synchrophasor} also attempted to find $\boldsymbol{\alpha}$ that minimizes the residue magnitude, but they assumed that $\|\boldsymbol{\alpha}\|_{0} = 1$,~\ie, only a single PMU is spoofed at a given time; this assumption simplifies the problem significantly, but it makes the approach in~\cite{fan2017synchrophasor} not applicable to the case where more than one PMU is being spoofed.

\begin{algorithm}[t]
\caption{Sparse GPS spoofing correction algorithm}
\textbf{Init.}: $\hat{\boldsymbol{\alpha}}=\textbf{0},~\textbf{r}^{[0]} = (\textbf{I}_{m} - \Pbf_{\Hbf})\boldsymbol{\Phi}^{-1}(\hat{\boldsymbol{\alpha}})\bar{\zbf}, \Amsc^{[0]} =\emptyset,~itr = 1$
\begin{algorithmic}[1]
    \WHILE{$\|\rbf^{[itr-1]}\|_2^2 > \tau $}
      \STATE Compute normalized residue magnitudes: 
      \begin{center}
          $\Tilde{r}_{i} = \frac{\|\rbf^{[itr-1]}_{i}\|_2^2}{m_i},~~\forall~i\in \Tmsc\setminus\Amsc^{[itr - 1]}$
      \end{center}
      \STATE Select the largest normalized residue magnitude:
      \begin{center}
          $i^* = \argmax\limits_{i\in \Tmsc\setminus\Amsc^{[itr - 1]}}~~\Tilde{r}_{i} $ 
      \end{center}
          \STATE Update the support of $\boldsymbol{\alpha}$:
          \begin{center}
          $\mathscr{A}^{[itr]} \leftarrow \mathscr{A}^{[itr-1]}\cup \{i^*\}$    
          \end{center}
          \STATE Compute the estimate $\hat{\boldsymbol{\alpha}}$:
          \begin{center}
$    \hat{\boldsymbol{\alpha}} = \argmin\limits_{\boldsymbol{\alpha}: supp(\boldsymbol{\alpha})\subseteq \Amsc^{[itr]}} \Vert(\textbf{I}_m - \Pbf_{\Hbf})\boldsymbol{\Phi}^{-1}(\boldsymbol{\alpha})\bar{\zbf}\Vert_2^2.$
          \end{center}
          \STATE Update the residual
          \begin{center}
$       {\rbf}^{[itr]} = (\textbf{I}_{m} - \Pbf_{\Hbf})\boldsymbol{\Phi}^{-1}(\hat{\boldsymbol{\alpha}})\bar{\zbf}$
          \end{center}
\STATE $itr = itr + 1$
    \ENDWHILE
\STATE \textbf{Data correction: } $\hat{\zbf} = \boldsymbol{\Phi}^{-1}(\hat{\boldsymbol{\alpha}})\bar{\zbf}$
\OUTPUT $\hat{\boldsymbol{\alpha}}$ and $\hat{\zbf}$
  \end{algorithmic}
  \label{alg: algorithm}
\end{algorithm}

We propose a greedy iterative algorithm to solve~\eqref{eq: optimization_problem} efficiently, which has a similar structure with existing residue-based greedy algorithms such as orthogonal matching pursuit~\cite{tropp2007signal}. The flow chart given in Figure~\ref{fig: Flow chart} illustrates the high-level operation of the sparse error correction algorithm and the detailed pseudocode is given in Algorithm~\ref{alg: algorithm}. As shown in Step 2 of Algorithm~\ref{alg: algorithm}, we first evaluate the normalized projection residue magnitudes using the projection residue~\eqref{eq: proj residue} computed based on the estimated set of spoofed PMUs and the attack vector estimate (from the previous iteration). Then we find the PMU with the largest normalized projection residue in Step 3 of Algorithm~\ref{alg: algorithm} and add that to the estimated set of spoofed PMU in Step 4. In Step 5 of Algorithm~\ref{alg: algorithm}, we update the estimate of the attack vector $\boldsymbol{\alpha}$ accordingly by solving a nonlinear least squares problem with a support constraint representing the estimated set of spoofed PMUs. In Step 6 of the algorithm, we update the projection residue based on the current estimate of the attack vector. The above steps are iterated until the squared projection residue magnitude becomes smaller than the preset threshold $\tau$.
Note that in Algorithm~\ref{alg: algorithm}, $\textbf{r}_i \in \mathbb{C}^{m_i}$ denotes the projection residue of measurements from the PMU at bus $i\in\Tmsc$.

\textbf{Operating on observable and unobservable systems:}
Unlike most of the existing data correction algorithms, the proposed algorithm can be used on both observable and unobservable systems. This is due to the fact that our formulation of attack angle bias estimation problem in~\eqref{eq: optimization_problem} aims at directly estimating $\boldsymbol{\alpha}$ without jointly estimating the state vector $\xbf$. It simply relies on the fact that with the true $\boldsymbol{\alpha}$, $\Phi^{-1}(\boldsymbol{\alpha})\bar{\zbf}$ should lie very close to $\Rmsc(\Hbf)$ and thus does not require observability of the system based on PMU measurements. This can be formally verified by our identifiability theorems, which explicitly guarantees that the attack can be identified as long as the per-zone conditions in Theorem~\ref{thm : condition on sparsity of alpha - local} are satisfied regardless of the state observability. We experimentally verify this in the Section~\ref{sec:exp} where we demonstrate the efficacy of the proposed algorithm under unobservable PMU placements of RTS-96 and IEEE-300 bus test networks.

~\textbf{Scalable implementation of Steps 2, 5, and 6:} Steps 2, 5 and 6 of Algorithm~\ref{alg: algorithm} are the most computationally heavy steps in each iteration. We exploit the decomposibility of the measurement model to solve these steps in a computationally efficient manner. From the decomposed measurement model~\eqref{eq: decoupled model} we can infer that $\boldsymbol{\Phi}^{-1}(\boldsymbol{\alpha})$ and $(\Ibf_m - \Pbf_{\Hbf})$ are block matrices, wherein off-diagonal blocks are equal to zero matrices and each diagonal block corresponds to a zone in the network, as shown below:
\begin{equation*}
\label{eq: decoupled Phi alpha}
\boldsymbol{\Phi}^{-1}(\boldsymbol{\alpha}) =\begin{bmatrix}
 \boldsymbol{\Phi}_1^{-1}(\boldsymbol{\alpha}^{(1)}) & \dots & \obf\\
 \vdots & \ddots &\vdots\\
 \obf & \dots & \boldsymbol{\Phi}_\Gamma^{-1}(\boldsymbol{\alpha}^{(\Gamma)}) 
 \end{bmatrix},~\textrm{and}
\end{equation*}
\begin{equation*}
\label{eq: decoupled orthogonal projection matrix}
 (\boldsymbol{I}_m - \Pbf_{\Hbf}) = \begin{bmatrix}
  (\boldsymbol{I}_{m^{(1)}} - \Pbf_{\Hbf^{(1)}}) & \dots & \obf\\
 \vdots & \ddots &\vdots\\
 \obf & \dots &  (\boldsymbol{I}_{m^{(\Gamma)}} - \Pbf_{\Hbf^{(\Gamma)}})
 \end{bmatrix}.
\end{equation*}
Hence Step 5 of Algorithm~\ref{alg: algorithm} is equivalent to,
\begin{equation}
\label{eq: residue minimization_2}
\resizebox{1\hsize}{!}{$
\begin{array}{ccc}
    \hat{\boldsymbol{\alpha}} & = & \argmin\limits_{\underset{supp(\boldsymbol{\alpha})\subseteq \Amsc^{[itr]}}{\boldsymbol{\alpha}:~}} \sum\limits_{\gamma=1}^{\Gamma} \Vert(\textbf{I}_{m^{(\gamma)}} - \Pbf_{\Hbf^{(\gamma)}})\boldsymbol{\Phi}_\gamma^{-1}(\boldsymbol{\alpha}^{(\gamma)})\bar{\zbf}^{(\gamma)})\Vert_2^2.\\
\end{array}$}
\end{equation}
This optimization can be solved independently per zone, \ie,
\begin{equation}
\label{eq: residue minimization_3}
\resizebox{1\hsize}{!}{$
\begin{array}{ccc}
    \hat{\boldsymbol{\alpha}}^{(\gamma)} & = &  \argmin\limits_{\underset{supp(\boldsymbol{\alpha}^{(\gamma)})\subseteq \Amsc^{(\gamma),[itr]}}{\boldsymbol{\alpha}^{(\gamma)}:}} \Vert(\textbf{I}_{m^{(\gamma)}} - \Pbf_{\Hbf^{(\gamma)}})\boldsymbol{\Phi}_\gamma^{-1}(\boldsymbol{\alpha}^{(\gamma)})\bar{\zbf}^{(\gamma)})\Vert_2^2,\\ 
\end{array}$}
\end{equation}
where $\Amsc^{(\gamma),[itr]}$ is the subset of the elements of $\Amsc^{[itr]}$ that belongs to Zone $\gamma$. Let $\gamma^*$ be the zone that contains bus $i^*$ selected in Step 3 of the algorithm. Since $\Amsc^{(\gamma),[itr]} = \Amsc^{(\gamma),[itr-1]}$ for $\gamma\neq \gamma^*$, $\hat{\boldsymbol{\alpha}}^{(\gamma)}$ at iteration $itr$ remains unchanged from the previous iteration for all $\gamma\neq \gamma^*$. The update on $\boldsymbol{\alpha}$ happens only at Zone $\gamma^*$, as given below:
\begin{equation}
\label{eq: residue minimization_4}
\resizebox{1\hsize}{!}{$
\begin{array}{ccc}
    \hat{\boldsymbol{\alpha}}^{(\gamma^*)} & = &  \argmin\limits_{\underset{supp(\boldsymbol{\alpha}^{(\gamma^*)})\subseteq \Amsc^{(\gamma^*),[itr]}}{\boldsymbol{\alpha}^{(\gamma^*)}:}} \Vert(\textbf{I}_{m^{(\gamma^*)}} - \Pbf_{\Hbf^{(\gamma*)}})\boldsymbol{\Phi}_{\gamma^*}^{-1}(\boldsymbol{\alpha}^{(\gamma^*)})\bar{\zbf}^{(\gamma^*)})\Vert_2^2,\\ 
\end{array}$}
\end{equation}
Therefore, Step 5 reduces to solving a least squares problem for only one zone,~$\gamma^*$. We employ gradient descent algorithm with backtracking line search to solve the above optimization problem.

Similarly, due to the special block structure of $\boldsymbol{\Phi}^{-1}(\boldsymbol{\alpha})$ and $(\Ibf_m - \Pbf_{\Hbf})$, the projection residue $\rbf^{[itr]}$ can be decomposed into zones. Then, since the only update of $\hat{\boldsymbol{\alpha}}$ happens at $\hat{\boldsymbol{\alpha}}^{(\gamma^*)}$, the residue update at Step 6 of Algorithm~\ref{alg: algorithm} simplifies to,
\begin{equation}
\label{eq: residue minimization_5}
\rbf^{(\gamma),[itr]} =
\begin{cases}
& (\Ibf_{m^{(\gamma^*)}} - \Pbf_{\Hbf^{(\gamma^*)}})\boldsymbol{\Phi}_{\gamma^*}^{-1}(\hat{\boldsymbol{\alpha}}^{(\gamma^*)})\bar{\zbf}^{(\gamma^*)}~~\gamma=\gamma^*\\
& \rbf^{(\gamma),[itr-1]}~~~~~~~~~~~~~~~~~~~~~~~~~~~~~\forall~\gamma\neq\gamma^*
\end{cases}
\end{equation}
\noindent where $\rbf^{(\gamma),[itr]}$ denotes the sub-vector of $\rbf^{[itr]}$ corresponding to Zone $\gamma$. 

Moreover, according to~\eqref{eq: residue minimization_5}, the only change in projection residue vector happens in the entries corresponding to zone $\gamma^*$. Therefore, in Step 2 of the algorithm, we can simply update the normalized residue magnitudes $\Tilde{r}_i$ for the PMUs $i$ that belong in Zone $\gamma^*$. Hence the computations in Steps 2, 5 and 6 of the algorithm are reduced to single-zone updates. This significantly reduces the computational complexity of the algorithm and makes it scalable.

\begin{table}[t]
\footnotesize
\caption{Asymptotic analysis of step-wise computational complexity of the algorithm per iteration}
\begin{tabular}{p{2 cm} p{6cm}}
Algorithm Step & Computational complexity \\
\hline \hline
Initialization & $O(m^2)$ \\
\hline
Step 2 & $O(K_{\max})$ \\
\hline
Step 3 & $O(K\log{K})$ \\
\hline
Step 4 & $O(1)$\\
\hline
Step 5 & $O((K_{\max})^2)$ per gradient-descent iteration \\
\hline
Step 6 & $O((K_{\max})^2)$ \\
\hline
\end{tabular}
\label{tab:compuational complexity}
\end{table}

In Table~\ref{tab:compuational complexity} we present the per-iteration complexity of the aforementioned scalable implementation of our sparse error correction algorithm. The major computation cost in the algorithm is due to Step 2, Step 5 and Step 6. The overall complexity of one iteration of gradient descent algorithm in Step 5, which involves computing the gradient and evaluating the objective function for the line search algorithm takes $O((K_{\max})^2)$, where $K_{\max}\triangleq{\max\limits_{\gamma=1,\dots,\Gamma}K^{(\gamma)}}$. This is due to the fact that the optimization problem~\eqref{eq: residue minimization_4} solved in Step 5 is quadratic in $e^{j\alpha_k}$ terms.
Furthermore, assuming that the degree of each bus in the power network is uniformly bounded, we can easily see that Step 2 requires $O(K_{\max})$ computation and Step 6 requires $O((K_{\max})^2)$ computation. Note that the computation time of each step depends only on $K_{\max}$ because Steps 2, 5, and 6 requires computation only for the single selected zone as described in our earlier discussion of the scalable implementation. This implies that the algorithm is scalable, therein the complexity of the algorithm is independent from the size of the power network and rather depends on the size of the largest zone in the network.

\section{Experiments}
\label{sec:exp}
In this section, we perform an extensive analysis of the efficacy of the proposed PMU data correction algorithm on the IEEE RTS-96 test network and IEEE-300 bus test network.


\begin{table}[t]
\parbox{1\linewidth}{
\footnotesize
\caption{Comparison of $\left(\textrm{median} \pm \frac{\textrm{standard deviation}}{2}\right)$, and the maximum of $\|\hat{\boldsymbol{\alpha}} - \boldsymbol{\alpha}\|_{\infty}$ from 100 Monte Carlo runs, for RTS-96 observable pmu placement setting, in degrees}
\begin{tabular}{p{1.1cm} p{2.1cm} p{2.1cm} p{2.1cm}}
\multirow{2}{1.2cm}{\makecell{Spoofed \\ PMU \%}}  & & \\
\cline{2-4}
& \makecell{~~~Proposed~~~~~} & \makecell{Risbud ~\emph{et al.} \\ \cite{risbud2018vulnerability}} & \makecell{Vanfretti~\emph{et al.} \\ \cite{vanfretti2010phasor}} \\
\hline \hline
$10\%$ & \makecell[l]{0.200 $\pm$ 0.165\\(Max.: 1.590)} & \makecell[l]{1.360 $\pm$ 0.819\\(Max.: 9.999)} & \makecell[l]{3.415 $\pm$ 1.166\\(Max.: 13.588)} \\
\hline
$20\%$ & \makecell[l]{0.580 $\pm$ 0.163\\(Max.: 1.353)} & \makecell[l]{4.393 $\pm$ 1.812\\ (Max.: 20.418)} & \makecell[l]{3.964 $\pm$ 1.359\\(Max.: 14.546)}  \\
\hline
$30\%$ & \makecell[l]{0.789 $\pm$ 0.165\\(Max.: 2.095)} & \makecell[l]{6.414 $\pm$ 2.158\\(Max.: 21. 504)} & \makecell[l]{3.733 $\pm$ 1.364\\(Max.: 12.755)}  \\
\hline
$40\%$ & \makecell[l]{0.853 $\pm$ 0.337\\(Max.: 1.990)} & \makecell[l]{6.634 $\pm$ 1.919\\(Max.: 18.928)} & \makecell[l]{3.164 $\pm$ 1.398\\(Max.: 13.693)}  \\
\hline
\end{tabular}
\label{tab:RTS-96 results observable}}

\parbox{1\linewidth}{
\footnotesize
\centering
\caption{Comparison of $\left(\textrm{median} \pm \frac{\textrm{standard deviation}}{2}\right)$, and the maximum of $\|\hat{\boldsymbol{\alpha}} - \boldsymbol{\alpha}\|_{\infty}$ from 100 Monte Carlo runs, for RTS-96 unobservable pmu placement setting, in degrees}
\label{tab:RTS-96 results unobservable}
\begin{tabular}{p{1.1cm} p{2.1cm} p{2.1cm}}
\multirow{2}{1.2cm}{\makecell{Spoofed \\ PMU \%}} &  &\\
\cline{2-3}
 & \makecell{~~~Proposed~~~~~} & \makecell{Vanfretti~\emph{et al.} \\ \cite{vanfretti2010phasor}} \\
\hline \hline 
$10\%$ & \makecell[l]{0.218 $\pm$ 0.147\\(Max.: 1.461)} & \makecell[l]{4.470 $\pm$ 1.967\\(Max.: 23.220)} \\
\hline
$20\%$ & \makecell[l]{0.703 $\pm$ 0.210\\(Max.: 2.133)} & \makecell[l]{5.691 $\pm$ 2.127\\ (Max.: 24.768)}\\
\hline
$30\%$ & \makecell[l]{0.678 $\pm$ 0.179\\(Max.: 1.839)} & \makecell[l]{4.511 $\pm$ 1.598\\ (Max.: 18.161)}\\
\hline
$40\%$ & \makecell[l]{0.809 $\pm$ 0.177\\(Max.: 1.867)} & \makecell[l]{4.389 $\pm$ 1.578\\(Max.: 16.366)} \\
\hline
\end{tabular}}
\end{table}


\textbf{Benchmark algorithms: }We compare the performance of the proposed approach with two existing benchmark algorithms, Risbud~\emph{et al.}~\cite{risbud2018vulnerability} and Vanfretti~\emph{et al.}~\cite{vanfretti2010phasor}. Risbud~\emph{et al.}~\cite{risbud2018vulnerability} presents an alternating minimization algorithm for joint state estimation and attack reconstruction. This algorithm is designed to operate on networks that are observable from the PMU measurements~\footnote{In~\cite{risbud2018vulnerability}, the authors extended their approach to make it applicable to a network that is not observable based on PMU measurements by incorporating SCADA measurements into their approach.  Since our focus here is on evaluating PMU data correction algorithms using only PMU measurements, their approach using both PMU and SCADA measurements is not considered in our comparative analysis.}. Vanfretti~\emph{et al.}~\cite{vanfretti2010phasor} develops a state estimation technique based on PMU measurements by incorporating potential phase bias errors in PMU measurements~\footnote{For RTS-96 network we set PMU 102 as the  "reference bus" defined in the paper~\cite{vanfretti2010phasor}, and for IEEE-300 network we set it to PMU 1. This PMU is assumed to be intact from PMU attacks.}. This algorithm is designed for decentralized operation wherein it can be independently applied to correct PMU data in observable islands within an unobservable network.  

As the performance metric, we employ the largest magnitude entry of the attack estimation error vector  $(\hat{\boldsymbol{\alpha}} - \boldsymbol{\alpha})$,~\ie, $\|\hat{\boldsymbol{\alpha}} - \boldsymbol{\alpha}\|_{\infty} = \max\limits_{i = \{1\dots,K\}} |\hat{\boldsymbol{\alpha}}_i - \boldsymbol{\alpha}_i|$. This metric measures the largest among the absolute angle bias estimation errors for all PMUs in the grid. Therefore it quantifies how good the worst performance of the data correction algorithm is. For each experiment we present the median, standard deviation and maximum of this performance metric over 100 Monte-Carlo runs.

\textbf{RTS-96 test network:} Here we demonstrate the efficacy of the proposed PMU data correction algorithm on the IEEE RTS-96 test network~\cite{grigg1999ieee}, which consists of 73 buses and 120 branches. We evaluate our data correction algorithm on both observable and unobservable PMU placements. Figure~\ref{fig:RTS96} illustrates the observable PMU placement setting which consists of 21 PMUs. Each deployed PMU measure the voltage phasor at the installed bus and the current phasors in all the branches incident to that bus. As described in Figure~\ref{fig:RTS96} this placement setting naturally induces two zones in the network, with 7 PMUs in Zone 1 and 14 PMUs in Zone 2. Furthermore, we obtained an unobservable PMU network by removing PMUs at buses 303, 103, and 316 in the observable placement, which results in a network with two zones, having 5 PMUs in Zone 1 and 13 PMUs in Zone 2. This causes around $15\%$ of the buses in the network to become unobservable. 

We test the proposed data correction algorithm on measurements generated according to~\eqref{eq: combined model} by sampling the state $\xbf$ from a Gaussian distribution with mean set to a known snapshot state and standard deviations of voltage magnitudes and phase angles set to 0.01 p.u. and 5.73 degrees, respectively. Given that the percentage of the spoofed PMUs is set to $A\%$, in each Monte Carlo run, we selected $A\%$ of PMUs from each zone uniformly at random and manipulated their phase angle measurements according to the spoofing attack model~\eqref{eq: combined model}. The magnitude of attack angle bias $\alpha_k$ for a spoofed PMU $k$ is sampled uniformly at random from the intervals in the range of $(-16^{\circ} , -24^{\circ}) \cup (16^{\circ} , 24^{\circ}).$ Furthermore, to accurately emulate the real world PMU measurements, we add Gaussian noise to both real and imaginary parts of the phasor measurements, with 0 mean and 0.01 standard deviation.


We first present the results for observable PMU placement setting shown in Figure~\ref{fig:RTS96}, where an equal percentage of PMUs are spoofed from Zone 1 and Zone 2. Table~\ref{tab:RTS-96 results observable} presents the median, standard deviation, and the maximum value of this metric from 100 Monte Carlo runs, for various percentages of the spoofed PMUs and for an observable PMU placement setting. The medians and the standard deviations indicate that the proposed sparse error correction approach significantly outperforms the benchmarks on average. In the meanwhile, the maximum error metrics observed among 100 Monte Carlo runs imply that our approach is more reliable compared to the benchmarks. For instance, the error metric remains smaller than 2.1 degrees for our approach in all Monte Carlo runs and all experiment scenarios, but for the benchmarks, the error metric can grow even larger than 12 degrees for some worst case attack scenarios. Table~\ref{tab:RTS-96 results unobservable} presents the results for mitigating GPS spoofing attacks carried out on an unobservable PMU network, where the rest of the attack is designed similarly to the experiments with the observable network. The results show a similar trend as the results for the observable case. In all of the above experiments the percentage of spoofed PMUs in each zone remains less than half of the number of PMUs in the zone. Therefore the attacks we test here are identifiable based on Theorem~\ref{thm : condition on sparsity of alpha - local} and thus the sparse error correction algorithm can identify and correct them well.

\begin{figure}[t]
 \centering
 \includegraphics[width=0.5\textwidth]{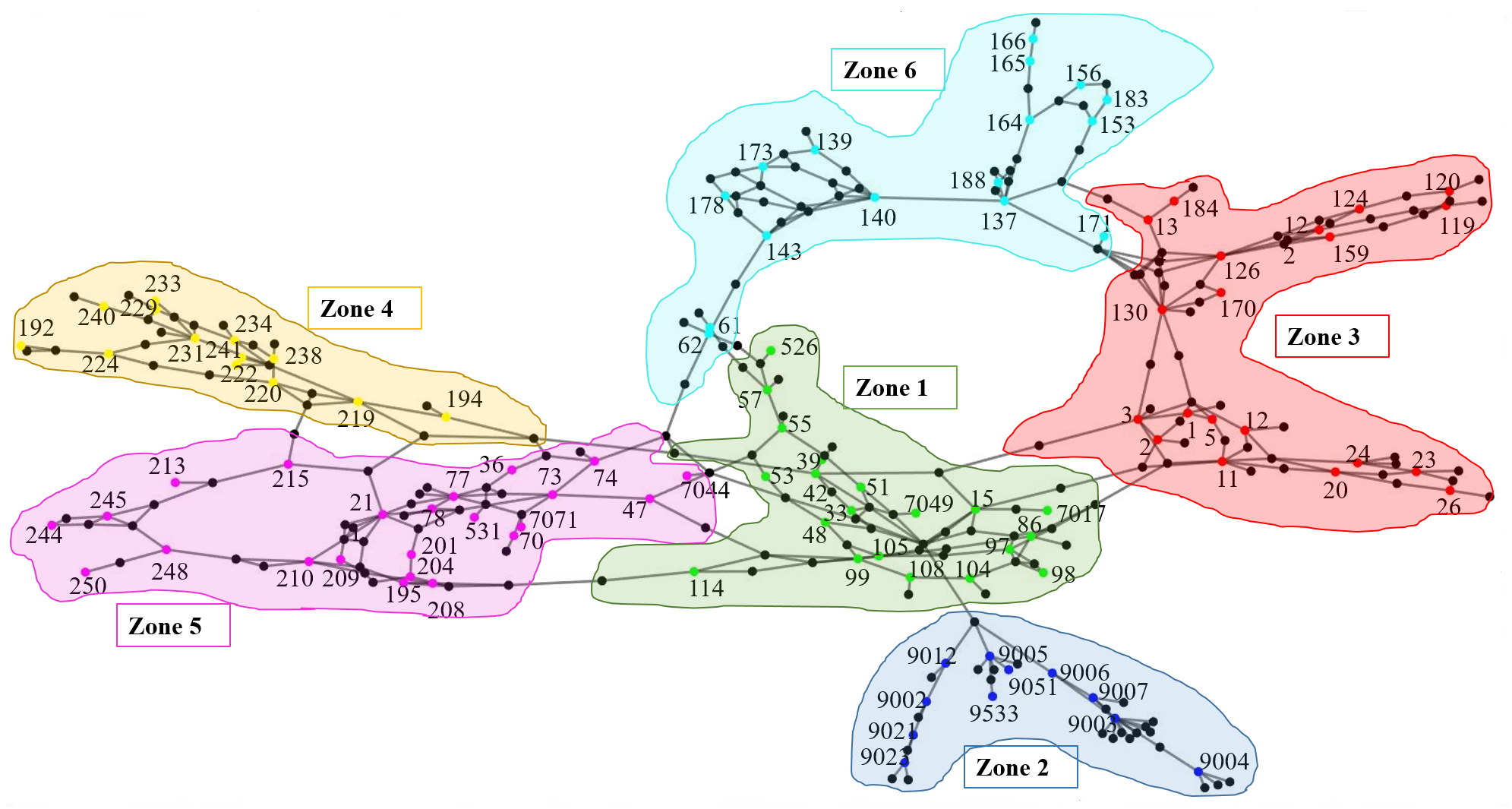}
 \caption{Observable PMU placement in IEEE-300 test network and the six zones induced by these PMUs. PMUs in each zone are indicated by colored dots, and their corresponding bus numbers.}
 \label{fig:IEEE300}
\end{figure}

\textbf{IEEE-300 test network:} IEEE-300 bus test network, consists of 300 buses and 411 branches. Figure~\ref{fig:IEEE300} illustrates the PMU locations of IEEE-300 bus network assumed in our experiment for the case that the network is assumed to be observable based on PMUs. As shown in Figure~\ref{fig:IEEE300}, this placement setting naturally induces six zones in the network, with 20, 11, 20, 13, 23 and 16 PMUs in Zone 1 to Zone 6 respectively. In addition we also perform experiments with an unobservable PMU placement. To obtain the unobservable PMU placement, we remove 13 PMUs~\footnote{The removed PMUs are: 26, 47, 70, 86, 114, 119, 159, 184, 213, 244, 526, 7017, and 7044.} from the observable placement, which results in 6 zones with 16, 11, 16, 13, 18, and 16 PMUs. The PMU measurement generation and attack implementation was performed in a way similar to the IEEE RTS-96 experiment.

Table~\ref{tab:IEEE-300 results observable} and Table~\ref{tab:IEEE-300 results unobservable} present the statistics of the performance metric $\|\hat{\boldsymbol{\alpha}} - \boldsymbol{\alpha}\|_{\infty}$, as the percentage of the spoofed PMUs increases from $10\%$ to $30\%$, for observable and unobservable PMU placements, respectively. Overall, the results follow a similar trend to the results we obtained from the RTS-96 case experiment. The benchmark approaches resulted in large errors in angle bias estimation, especially when the percentage of spoofed PMUs is large. On the other hand, the proposed sparse error correction approach consistently showed small medians of error metrics demonstrating its effectiveness in mitigating the spoofing attacks. As shown in Table~\ref{tab:IEEE-300 results observable}, for observable PMU placement setting, the proposed approach missed the detection of some spoofed PMUs in 2 out of 100 Monte Carlo runs when 30\% of PMUs are under spoofing attacks and resulted in a large error; however, in other MC runs, the approach successfully localized spoofed PMUs and resulted in accurate estimates of the angle biases. From Table~\ref{tab:IEEE-300 results unobservable} we observe that detecting the attacks in the unobservable PMU placement setting is comparatively more challenging to the proposed algorithm. For instance when $30\%$ of PMUs are under spoofing attack, 4 out of 100 Monte Carlo runs report high maximum angle estimation errors due to improper detection of some spoofed PMUs in these Monte Carlo runs.
 
 \begin{table}[t]
\parbox{1\linewidth}{
\footnotesize
\caption{Comparison of $\left(\textrm{median} \pm \frac{\textrm{standard deviation}}{2}\right)$, and the maximum of $\|\hat{\boldsymbol{\alpha}} - \boldsymbol{\alpha}\|_{\infty}$ from 100 Monte carlo runs, for IEEE-300 observable pmu placement setting, in degrees}
\label{tab:IEEE-300 results observable}
\begin{tabular}{p{1.1cm} p{2.1cm} p{2.1cm} p{2.1cm}}
\multirow{2}{1.2cm}{\makecell{Spoofed \\ PMU \%}}  & & \\
\cline{2-4}
& \makecell{~~~Proposed~~~~~} & \makecell{Risbud ~\emph{et al.} \\ \cite{risbud2018vulnerability}} & \makecell{Vanfretti~\emph{et al.} \\ \cite{vanfretti2010phasor}} \\
\hline \hline
$10\%$ & \makecell[l]{1.185 $\pm$ 0.213\\(Max.: 2.455)} & \makecell[l]{ 6. 682 $\pm$ 2.625 \\(Max.: 23.8574)} & \makecell[l]{ 17.590 $\pm$ 5.786 \\(Max.: 80.920)} \\
\hline
$20\%$ & \makecell[l]{ 1.288$\pm$ 0.298\\(Max.: 4.0146)} & \makecell[l]{ 10.331$\pm$ 3.129 \\ (Max.: 26.619 )} & \makecell[l]{ 15.138 $\pm$ 4.753\\(Max.: 54.985 )}  \\
\hline
$30\%$ & \makecell[l]{ 1.542 $\pm$ 1.630 \\(Max.: 22.756)} & \makecell[l]{ 15.764$\pm$2.754 \\(Max.: 25.850 )} & \makecell[l]{17.549 $\pm$ 5.484 \\(Max.: 76.824)}  \\
\hline
\end{tabular}}

\parbox{1\linewidth}{
\footnotesize
\centering
\caption{Comparison of $\left(\textrm{median} \pm \frac{\textrm{standard deviation}}{2}\right)$, and the maximum of $\|\hat{\boldsymbol{\alpha}} - \boldsymbol{\alpha}\|_{\infty}$ from 100 Monte Carlo runs, for IEEE-300 unobservable pmu placement setting, in degrees}
\label{tab:IEEE-300 results unobservable}
\begin{tabular}{p{1.1cm} p{2.1cm} p{2.1cm}}
\multirow{2}{1.2cm}{\makecell{Spoofed \\ PMU \%}} &  &\\
\cline{2-3}
 & \makecell{~~~Proposed~~~~~} & \makecell{Vanfretti~\emph{et al.} \\ \cite{vanfretti2010phasor}} \\
\hline \hline 
$10\%$ & \makecell[l]{0.820 $\pm$ 1.185\\(Max.: 24.300)} & \makecell[l]{17.989 $\pm$ 5.782\\(Max.: 74.151} \\
\hline
$20\%$ & \makecell[l]{1.310 $\pm$ 1.160\\(Max.: 23.978)} & \makecell[l]{17.721 $\pm$ 4.720\\ (Max.: 25.162)}\\
\hline
$30\%$ & \makecell[l]{1.420 $\pm$ 1.582\\(Max.: 22.677)} & \makecell[l]{19.235 $\pm$ 4.312\\ (Max.: 49.085)}\\
\hline
\end{tabular}}
\end{table}

\begin{table*}[t]
\parbox{1\linewidth}{
\footnotesize
\centering
\caption{Performance of the proposed algorithm under ramping attack on IEEE RTS-96 observable PMU network. The performance statistics presented are $\left(\textrm{median} \pm \frac{\textrm{standard deviation}}{2}\right)$, and the maximum of $\|\hat{\boldsymbol{\alpha}}^{[t]} - \boldsymbol{\alpha}^{[t]}\|_{\infty}$ at sample point $t$, in degrees.}
\label{tab:ramping attack}
\begin{tabular}{p{1.1cm} p{2.1cm} p{2.1cm} p{2.1cm}p{2.1cm} p{2.1cm} p{2.1cm}}
\multirow{2}{1.2cm}{\makecell{Spoofed \\ PMU \%}}  & \multicolumn{6}{c}{$\frac{\textrm{Attack angle bias}}{\textrm{Maximum angle bias}}\times 100\%$} \\
\cline{2-7}
& \makecell{~~~0\%~~~~~} & \makecell{~~~20\%~~~~~} & \makecell{~~~40\%~~~~~} & \makecell{~~~60\%~~~~~} & \makecell{~~~80\%~~~~~} & \makecell{~~~100\%~~~~~} \\
\hline \hline
$10\%$ & \makecell[l]{0 $\pm$ 0.033\\(Max.: 0.658)} & \makecell[l]{  0.187$\pm$ 0.086 \\(Max.: 0.798)} & \makecell[l]{ 0.177$\pm$0.107 \\(Max.:0.911 )}  & \makecell[l]{ 0.198$\pm$0.089 \\(Max.:0.908 )}  & \makecell[l]{ 0.183$\pm$0.099 \\(Max.:0.917 )} & \makecell[l]{ 0.195$\pm$0.091 \\(Max.:0.860 )} \\
\hline
$20\%$ & \makecell[l]{ 0$\pm$ 0.076\\(Max.: 0.919)} & \makecell[l]{ 0.353$\pm$0.095 \\ (Max.:1.100 )} & \makecell[l]{ 0.364$\pm$0.086\\(Max.:0.932 )}  & \makecell[l]{ 0.381$\pm$0.101 \\(Max.:0.998)}  & \makecell[l]{ 0.394$\pm$0.083 \\(Max.:0.817)}  & \makecell[l]{ 0.382$\pm$0.097 \\(Max.:0.925)} \\
\hline
$30\%$ & \makecell[l]{ 0 $\pm$ 0.068 \\(Max.: 0.884)} & \makecell[l]{ 0.458$\pm$0.094 \\(Max.:1.109 )} & \makecell[l]{ 0.466$\pm$0.077 \\(Max.:0.955 )}  & \makecell[l]{ 0.471$\pm$0.077 \\(Max.:0.945 )}  & \makecell[l]{ 0.499$\pm$0.081 \\(Max.:1.078 )}  & \makecell[l]{ 0.470$\pm$0.100 \\(Max.:1.130 )}  \\
\hline
$40\%$ & \makecell[l]{ 0 $\pm$ 0.060 \\(Max.: 0.868)} & \makecell[l]{ 0.493$\pm$0.087 \\(Max.:1.097)} & \makecell[l]{0.507$\pm$0.084 \\(Max.:0.965 )}  & \makecell[l]{ 0.495$\pm$0.094 \\(Max.:1.196 )}  & \makecell[l]{ 0.511$\pm$0.078 \\(Max.:1.006 )}  & \makecell[l]{ 0.476$\pm$0.076 \\(Max.:0.997 )} \\
\hline
\end{tabular}}
\end{table*}

\textbf{Performance under ramping attack:} We extend our experiments to present the effectiveness of the proposed algorithm to defend spoofing attacks designed as ramping attacks. In the ramping attack the attack angle bias of a spoofed PMU continues to linearly ramp up from 0 to a maximum angle bias value throughout the attack interval. The maximum angle bias is sampled uniformly at random from the intervals in the range of $(-16^{\circ} , -24^{\circ}) \cup (16^{\circ} , 24^{\circ})$. We test the performance of the proposed algorithm on observable IEEE RTS-96 PMU network to defend ramping attacks. To evaluate the performance, we pick 6 sample points within the ramping attack interval where the attack magnitude reaches 0\%, 20\%, 40\%, 60\%, 80\%, and 100\% of the maximum angle bias. This allows us evaluate the performance at different stages of the ramping attack. Selection of the location of spoofed PMUs is done similar to the earlier experiments, where we choose equal percentage of spoofed PMUs from each zone uniformly at random. For the measurement noise, we added Gaussian noise with 0 mean and 0.005 standard deviation to the real and imaginary parts of the PMU measurements. In Table~\ref{tab:ramping attack} we compare the performance statistics of $\|\hat{\boldsymbol{\alpha}}^{[t]} - \boldsymbol{\alpha}^{[t]}\|_{\infty}$ from 100 Monte-Carlo runs, where $\boldsymbol{\alpha}^{[t]}$ denotes the true attack angle bias at sample point $t$ and $\hat{\boldsymbol{\alpha}}^{[t]}$ denotes the estimated attack angle bias at sample point $t$. The results are presented for the 6 sample points evaluated in the attack, when various percentages of PMUs are spoofed in the network. The performance statistics imply that the proposed algorithm can accurately estimate angle biases introduced by the spoofing attacks with gradually increasing attack magnitudes.

\textbf{Computation time} All the simulations are conducted using MATLAB 2018b on a machine with an Intel Xeon E3 processor and a 16 GB RAM. In Table~\ref{tab:simulation time} we present the average runtime of the proposed and benchmark algorithms, based on simulations performed on the observable PMU placement settings of both IEEE RTS-96 and IEEE 300 bus test networks, when 7 PMUs in the grid are under spoofing attacks. We use the lsqnonlin function~\footnote{This function implements the Levenberg-Marquardt algorithm~\cite{matlab_lsqnonlin}.} in Matlab to solve the non-linear weighted least square problem in~Vanfretti~\emph{et.al.}~\cite{vanfretti2010phasor}. The simulation times in the Table~\ref{tab:simulation time} further support the scalability aspect of the proposed algorithm as highlighted in Section~\ref{sec: algorithm}. Compared to IEEE RTS-96 PMU placement, the average zone size of IEEE 300 bus PMU placement increases roughly by a factor of 1.6. According to the computation time analysis in Section~\ref{sec: algorithm}, this zone size increase would cause the computation time of Step 5 and Step 6 of the proposed algorithm to increase by a factor of $(1.6)^2$. Further, while we used 1000 as the maximum number of iterations threshold of the gradient descent algorithm for Step 5 in the IEEE RTS-96 test case, we increased this threshold to 4000 for the IEEE 300 bus case to ensure convergence since we need to solve a least squares problem with more variables. This is the reason behind the runtime increase of the proposed approach in IEEE 300 compared to IEEE RTS-96. As the grid size increases from IEEE RTS-96 to IEEE 300, the runtime of our approach scales significantly better than that of the benchmarks.

\begin{table}
\footnotesize
\caption{Average simulation time based on experiments with observable PMU placement setting of IEEE RTS-96 and IEEE 300 bus test networks.}
\begin{tabular}{p{1.8cm} p{1.8cm} p{1.8cm} p{1.8cm}}
\multirow{2}{1.2cm}{\makecell{ Network\\ }}  & & \\
\cline{2-4}
& \makecell{~~~Proposed~~~~~} & \makecell{Risbud ~\emph{et al.} \\ \cite{risbud2018vulnerability}} & \makecell{Vanfretti~\emph{et al.} \\ \cite{vanfretti2010phasor}} \\
\hline \hline
RTS-96 & \makecell[l]{0.338 sec} & \makecell[l]{~~~2.317 sec} & \makecell[l]{~~~~1.992 sec} \\
\hline
IEEE-300 & \makecell[l]{7.302 sec} & \makecell[l]{160.311 sec} & \makecell[l]{2028.781 sec}  \\
\hline
\end{tabular}
\label{tab:simulation time}
\end{table}

\section{Conclusion}
In this paper, we presented a sparse error correction framework for mitigating GPS spoofing attacks on PMUs. Our attack identifiability analysis provides a detailed characterization of how PMU locations affect the grid resilience to spoofing attacks. The proposed error correction algorithm is scalable because it only requires solving a least squares problem for a single zone in each iteration. It outperformed benchmarks in mitigating GPS spoofing attacks on PMUs. Overall, our results imply that we can mitigate spoofing attacks much more effectively by properly leveraging their sparse nature.\\\\

\noindent\textbf{Disclaimer: } This report was prepared as an account of work sponsored by an agency of the United States Government. Neither the United States Government nor any agency thereof, nor any of their employees, makes any warranty, express or implied, or assumes any legal liability or responsibility for the accuracy, completeness, or usefulness of any information, apparatus, product, or process disclosed, or represents that its use would not infringe privately owned rights. Reference herein to any specific commercial product, process, or service by trade name, trademark, manufacturer, or otherwise does not necessarily constitute or imply its endorsement, recommendation, or favoring by the United States Government or any agency thereof. The views and opinions of authors expressed herein do not necessarily state or reflect those of the United States Government or any agency thereof.

\bibliographystyle{IEEEtran}
\bibliography{references}

\appendices
\section{Alternative measurement model}
\label{app:alt_measurement}
In this section we present the proof for Proposition~\ref{prop : alternative measurements mapping} stated in Section~\ref{sec: identifiability}

First we will show that the spoofed voltage angle measurements of PMU in bus $i$ denoted by $\bar{w}_{\angle V_{i}}$ can be directly obtained from $\bar{\zbf}$, using the definition of  $\bar{w}_{\angle V_{i}}$, shown as below:
\begin{equation}  
\begin{aligned}
\bar{w}_{\angle V_{i}} & = \angle \bar{z}_{V_i}.
\end{aligned}
\end{equation}

Next we will show that the angle difference of voltage phase angles between bus $i\in\Tmsc$ and bus $l\in\Mmsc_i$ denoted by $\bar{w}_{\Delta \theta_{il}}$ can be derived using $\bar{\zbf}$.

Using~\eqref{eq: voltage phasor} and~(\ref{eq: spoofed voltage phasor}a), we can  model the spoofed voltage measurement from PMU $k$ in bus $i$ as shown below:
\begin{equation}  
\label{eq: spoofed voltage}
\begin{aligned}
\bar{z}_{V_i} & = e^{j\alpha_k}(|x_i|e^{j\theta_i}) = |x_i|e^{j(\alpha_k + \theta_i)}.
\end{aligned}
\end{equation}
Note that $\alpha_k$ is nonzero if PMU $k$ is under spoofing attack and $\alpha_k$ is zero otherwise.

Using~\eqref{eq: current phasor} and~(\ref{eq: spoofed voltage phasor}b), we can write down the current measurements in line from bus $i\in\Tmsc$ to bus $l\in\Mmsc_i$ as below,
\begin{equation}
\label{eq: spoofed current}
\begin{aligned}
\bar{z}_{I_{il}} =& e^{j\alpha_k}((y_{il} + j\frac{b^s_{il}}{2})|x_i|e^{j\theta_i} - y_{il}|x_l|e^{j\theta_l})\\
    =&  (y_{il} + j\frac{b^s_{il}}{2})|x_i|e^{j(\theta_i + \alpha_k)} - y_{il}|x_l|e^{j(\theta_l + \alpha_k)}.
\end{aligned}
\end{equation}
Now using~\eqref{eq: spoofed voltage} and~\eqref{eq: spoofed current}, we can write $\bar{z}_{V_i}\bar{z}_{I_{il}}^*$ as below:
\begin{equation}  
\label{eq: power flow}
\begin{aligned}
\bar{z}_{V_i}\bar{z}_{I_{il}}^* & =  |x_i|e^{j( \alpha_k + \theta_i)}(y_{il}^* - j\frac{b^s_{il}}{2})|x_i|e^{-j(\theta_i + \alpha_k)} \\ &~~~~~~~~~~ - |x_i|e^{j( \alpha_k + \theta_i)}y_{il}^*|x_l|e^{-j(\theta_l + \alpha_k)})\\
     & =  (y_{il}^* - j\frac{b^s_{il}}{2})|x_i|^2 - y_{il}^*|x_i||x_l|e^{j(\theta_i - \theta_l)}). \\
|x_l|e^{j(\theta_i - \theta_l)}) & = \frac {(y_{il}^* - j\frac{b^s_{il}}{2})|x_i|^2 - \bar{z}_{V_i}\bar{z}_{I_{il}}^* }{y_{il}^*|x_i|} \\
     \end{aligned}
\end{equation}
Since $|x_i|$ is directly measured from PMU at bus $i$, $|x_i| = |z_{V_i}|$. So we can rewrite the above euqation as
\begin{equation}  
\label{eq: power flow2}
\begin{aligned}
|x_l|e^{j(\theta_i - \theta_l)}) & = \frac {(y_{il}^* - j\frac{b^s_{il}}{2})|z_{V_i}|^2 - \bar{z}_{V_i}\bar{z}_{I_{il}}^* }{y_{il}^*|z_{V_i}|} \\
\end{aligned}
\end{equation}
Now we can write $\theta_i - \theta_l$, the angle difference of voltage phasors between bus  $i$ and bus $l$ denoted by $\bar{w}_{\Delta \theta_{il}}$ as
\begin{equation}
\label{eq: angle difference}
\begin{aligned}
    \bar{w}_{\Delta \theta_{il}} & = \theta_i - \theta_l = \angle (\frac{ (y_{il}^* - j\frac{b^s_{il}}{2})|z_{V_i}|^2 - \bar{z}_{V_i}\bar{z}_{I_{il}}^* }{y_{il}^*})\\
\end{aligned}
\end{equation}
Thus we prove that there exists a mapping $T(\cdot)$ such that $\bar{\wbf} = T(\bar{\zbf})$, where $\bar{\wbf}$ is the concatenation of $\bar{w}_{\angle V_i}$ and $\bar{w}_{\Delta \theta_{il}}$ for all $i\in\Tmsc$ and $l\in\Mmsc_i$. 

\section{Identifiability analysis}
\label{app : global identifiability definition}
In this section, we present proofs for the theorems, lemmas and propositions we have presented on identifiability analysis in Section~\ref{sec: identifiability}.
\subsection{Proof for Lemma~\ref{lemma : global cospark}}
\label{proof: lemma : global cospark}
\begin{proof}
Suppose $\boldsymbol{\alpha}$ satisfies
$
\|\boldsymbol{\alpha}\|_0 < \frac{1}{2}~\textrm{Cospark}(\Hbf_{\angle V}\Bbf_{\Delta})$ and that $\boldsymbol{\alpha}$ is not identifiable. Then by Proposition~\ref{prop : definition global identifiability}, there exists $\bar{\boldsymbol{\alpha}}\neq\boldsymbol{\alpha}$ such that $\|\bar{\boldsymbol{\alpha}}\|_0 \leq \|\boldsymbol{\alpha}\|_0$ and,
    \begin{equation}
    \label{eq: sufficient conditions from prop}
\boldsymbol{\alpha} - \bar{\boldsymbol{\alpha}} \in \Rmsc(\Hbf_{\angle V}\Bbf_{\Delta}).    
    \end{equation}
This implies that,
\begin{equation*}
    \begin{array}{ccc}
 \|\boldsymbol{\alpha} - \bar{\boldsymbol{\alpha}}\|_0 & \leq & \|\boldsymbol{\alpha}\|_0 +    \|\bar{\boldsymbol{\alpha}}\|_0 \\
 & \leq & 2\|\boldsymbol{\alpha}\|_0 \\
& <
&~\textrm{Cospark}(\Hbf_{\angle V}\Bbf_{\Delta})
    \end{array}
\end{equation*}

However, since $\boldsymbol{\alpha}-\bar{\boldsymbol{\alpha}}
\in \Rmsc{(\Hbf_{\angle V}\Bbf_{\Delta})}$ we have $ \|\boldsymbol{\alpha} - \bar{\boldsymbol{\alpha}}\|_0~\geq~\textrm{Cospark}(\Hbf_{\angle V}\Bbf_{\Delta})$, which contradicts with the above inequality.
Therefore, $\boldsymbol{\alpha}$ should be identifiable.
\end{proof}
\subsection{Proof of Proposition~\ref{prop: bdelta}}
\label{proof: prop: bdelta}
\begin{proof}
Since $\Hbf_\Delta\in\mbbR^{(m - K)\times N}$ is a block matrix with the special block structure as follows,
\begin{equation*}
\resizebox{0.5\hsize}{!}{$
     \Hbf_{\Delta} =
    \begin{bmatrix}
    \Hbf_{\Delta}^{(1)} & 0 & \hdots 0 \\
    0 & \Hbf_{\Delta}^{(2)} & \hdots  0  \\
    \vdots & \hdots&  \vdots\\
    0 & 0 & \hdots \Hbf_{\Delta}^{(\Gamma)}  
    \end{bmatrix},$}
\end{equation*}
we can describe its null space as:
\begin{equation}
\resizebox{0.75\hsize}{!}{$
\label{eq: null space of Hdelta}
    \begin{array}{ccc}
    \Nmsc({\Hbf_\Delta})  &=& \{\boldsymbol{\beta}\in\mbbR^{N}:\Hbf_\Delta \boldsymbol{\beta} = 0\}\\
    & =&  \left\{\begin{bmatrix}
        \boldsymbol{\beta}^{(1)}\\
        \vdots\\
        \boldsymbol{\beta}^{(\Gamma)}\\
    \end{bmatrix}\in\mbbR^{N}:\Hbf_\Delta^{(\gamma)} \boldsymbol{\beta}^{(\gamma)} = 0,~\forall \gamma\right\}
\end{array}\\$}
\end{equation}
Above equation implies that, 
\begin{equation}
    \boldsymbol{\beta}\in\Nmsc({\Hbf_\Delta})~\iff~          \boldsymbol{\beta}^{(\gamma)} \in \Nmsc(\Hbf_\Delta^{(\gamma)}),~\forall \gamma\in\{1,\dots,\Gamma\}
\end{equation}
Therefore, we can derive a basis matrix of $\Nmsc(\Hbf_\Delta)$ denoted by $\Bbf_\Delta$ with a similar block structure, having zero off diagonal blocks and $\gamma$-th diagonal block $B_\Delta^{(\gamma)}$ being the basis matrix of $\Nmsc(\Hbf_\Delta^{(\gamma)})$, as shown below:
\begin{equation*}
     \Bbf_{\Delta} =
    \begin{bmatrix}
    \Bbf_{\Delta}^{(1)} & 0 & \hdots 0 \\
    0 & \Bbf_{\Delta}^{(2)} & \hdots  0  \\
    \vdots & \hdots&  \vdots\\
    0 & 0 & \hdots \Bbf_{\Delta}^{(\Gamma)}  
    \end{bmatrix}.
\end{equation*}

Now let us prove that $B_\Delta^{(\gamma)}$ is the $N^{(\gamma)}$- dimensional vector with all entries equal to one. Recall that $\Hbf_\Delta^{(\gamma)} \in \mbbR^{(m^{(\gamma)}-K^{(\gamma)})\times N^{(\gamma)}}$ where $m^{(\gamma)}, K^{(\gamma)}$ and $N^{(\gamma)}$ are the number of measurements from Zone $\gamma$, number of PMUs in Zone $\gamma$ and number of buses in Zone $\gamma$, respectively. Suppose that $\bbf\in\mbbR^{N^{(\gamma)}}$ is a vector in $\Nmsc(\Hbf_{\Delta}^{(\gamma)})$, \ie,
\begin{center}
    $\Hbf_\Delta^{(\gamma)} \bbf = \boldsymbol{0}$
\end{center}
According to the definition, each row of $\Hbf_\Delta^{(\gamma)}$ consists of exactly two nonzero entries which are equal in magnitude but opposite in sign. Each column corresponds to a bus in Zone $\gamma$. Let the nonzero entries of row $i$ of  $\Hbf_\Delta^{(\gamma)}$ are located in column $k$ and column $l$. Then for each row $i$ of the matrix,
\begin{equation}
\label{eq: null space - zone z}
    \Hbf_\Delta^{(\gamma)}[i,:]\bbf = 0, \bbf\neq 0 \iff b_k = b_l,
\end{equation}
where $\Hbf_\Delta^{(\gamma)}[i,:]$ denotes the row $i$ of matrix $\Hbf_\Delta$ and $b_k$ denotes the entry $k$ of vector $\bbf$. 
Furthermore, from the definition of a zone, there exists a path between any two nodes in the subgraph corresponding to a particular zone. This implies that,~\eqref{eq: null space - zone z} has to be true for any entry $k$ and $l$ in the vector $\bbf$. Therefore $\bbf = \bar{b}\cdot\boldsymbol{1}_{N^{(\gamma)}}$ where $\bar{b}$ is a scalar. Thus we prove that $\Nmsc(\Hbf_\Delta^{(\gamma)})$ has dimension one and the basis for the null space is in fact the following: 
\begin{center}
    $\Bbf_\Delta^{(\gamma)} = \boldsymbol{1}_{N^{(\gamma)}}$.
\end{center}
\end{proof}
\subsection{Proof of Lemma~\ref{lemma : cospark in terms of Kmin}}
\label{proof: lemma : cospark in terms of Kmin}
\begin{proof}
Since both $\Hbf_{\angle V}$ and $\Bbf_{\Delta}$ are block matrices with off diagonal blocks being zero matrices, $\Hbf_{\angle V}\Bbf_{\Delta}$ is also a block matrix as shown below:
\begin{equation}
\label{eq: HvBp block diagonal matrix}
\resizebox{0.9\hsize}{!}{
$\begin{array}{lll}
    \Hbf_{\angle V}\Bbf_{\Delta} 
& = &
    \begin{bmatrix}
    \Hbf_{\angle V}^{(1)} & 0 & \hdots 0 \\
    0 & \Hbf_{\angle V}^{(2)} & \hdots  0  \\
    \vdots & \hdots&  \vdots\\
    0 & 0 & \hdots \Hbf_{\angle V}^{(\Gamma)}  \\
    \end{bmatrix}
    \begin{bmatrix}
    \Bbf_{\Delta}^{(1)} & 0 & \hdots 0 \\
    0 & \Bbf_{\Delta}^{(2)} & \hdots  0  \\
    \vdots & \hdots& \vdots\\
    0 & 0  & \hdots \Bbf_{\Delta}^{(\Gamma)}  \\
\end{bmatrix}\\\\
 & = & 
    \begin{bmatrix}
    \Hbf_{\angle V}^{(1)}\Bbf_{\Delta}^{(1)} & 0 & \hdots 0 \\
    0 & \Hbf_{\angle V}^{(2)}\Bbf_{\Delta}^{(2)} & \hdots  0  \\
    \vdots & \hdots&  \vdots\\
    0 & 0 & \hdots \Hbf_{\angle V}^{(\Gamma)}\Bbf_{\Delta}^{(\Gamma)}  
    \end{bmatrix},
\end{array}$
}
\end{equation}
where $\Hbf_{\angle V}^{(\gamma)}$ and $\Bbf_{\Delta}^{(\gamma)}$ are the blocks of $\Hbf_{\angle V}$ and $\Bbf_{\Delta}$ corresponding to each zone $\gamma\in\{1,2,\dots, \Gamma\}$. Now it is easy to see that the sparsity level of the sparsest nonzero vector in $\Rmsc({\Hbf_{\angle V}\Bbf_{\Delta}})$ is the smallest sparsity level among the sparsest nonzero vectors in $\Rmsc(\Hbf_{\angle V}^{(\gamma)}\Bbf_{\Delta}^{(\gamma)})$ of all zones $\gamma\in\{1,\dots,\Gamma\}$. Therefore by the definition of cospark of a matrix, we can prove the statement in the Lemma, \ie,
\begin{center}
    Cospark $(\Hbf_{\angle V}\Bbf_{\Delta})$ = $\min\limits_{\gamma\in \{1,2,\dots,\Gamma\}}\textrm{Cospark}(\Hbf_{\angle V}^{(\gamma)}\Bbf_{\Delta}^{(\gamma)})$.
\end{center}
\end{proof}
\subsection{Proof of Lemma~\ref{lemma : cospark in terms of Kmin2}}
\label{proof: lemma : cospark in terms of Kmin2}
\begin{proof}
We directly prove the claim in this Lemma by leveraging Proposition~\ref{prop: bdelta}
and the special sparsity structure of $\Hbf_{\angle V}^{(\gamma)}\in\mathbb{R}^{K^{(\gamma)}\times N^{(\gamma)}}$.
If $\ybf$ is a nonzero vector in $\Rmsc(\Hbf_{\angle V}^{(\gamma)} \Bbf_{\Delta}^{(\gamma)})$, then there exists a nonzero $c\in \mathbb{R}$ such that,
\begin{equation}
\label{eq: lemma III.3 intermediate}
\begin{array}{ccc}
\ybf & = & \Hbf_{\angle V}^{(\gamma)} \Bbf_{\Delta}^{(\gamma)}c\\
 & = & \Hbf_{\angle V}^{(\gamma)} \boldsymbol{1}_{N^{(\gamma)}}c~,\\
\end{array}
\end{equation}
where the last equation is due to Proposition~\ref{prop: bdelta}.
Furthermore, by the definition of $\Hbf_{\angle V}^{(\gamma)}$, each row of this matrix corresponds to a particular voltage angle measurement from a PMU deployed in bus $i\in\Tmsc^{(\gamma)}$ from Zone $\gamma$, and consists of all zeros except for the value one in the column corresponding to bus $i$. Due to this structure of $\Hbf_{\angle V}^{(\gamma)}$, we can say the following:
\begin{center}
    $\Hbf_{\angle V}^{(\gamma)}\boldsymbol{1}_{N^{(\gamma)}} = \boldsymbol{1}_{K^{(\gamma)}}$,
\end{center}
where $K^{(\gamma)}$ is the number of PMUs in Zone $\gamma$. Thus by substituting this in~\eqref{eq: lemma III.3 intermediate} we get,
\begin{equation*}
\begin{array}{ccc}
\ybf & = & \boldsymbol{1}_{K^{(\gamma)}}c\\
\end{array}
\end{equation*}
This implies that for any nonzero $\ybf\in \Rmsc(\Hbf_{\angle V}^{(\gamma)} \Bbf_{\Delta}^{(\gamma)})$, $\|\ybf\|_0 = K^{(\gamma)}$
Hence, from the definition of cospark,
\begin{equation*}
    \textrm{Cospark}(\Hbf_{\angle V}^{(\gamma)}\Bbf_{\Delta}^{(\gamma)}) = K^{(\gamma)}.
\end{equation*}
\end{proof}
\subsection{Proof of Theorem~\ref{thm : condition on sparsity of alpha - global}}
\label{proof: thm : condition on sparsity of alpha - global}
\begin{proof}
First we directly prove the implications of the inequality in the Theorem by leveraging  Lemma~\ref{lemma : global cospark}, Lemma~\ref{lemma : cospark in terms of Kmin} and Lemma~\ref{lemma : cospark in terms of Kmin2}.
Suppose that $\boldsymbol{\alpha}$ satisfies the following:
\begin{center}
$\begin{array}{ccc}
\|\boldsymbol{\alpha}\|_0 &<& \frac{1}{2}\textrm{Cospark}(\Hbf_{\angle V}\Bbf_{\Delta})\\ 
\end{array}$
\end{center}
Then from Lemma~\ref{lemma : global cospark}, $\boldsymbol{\alpha}$ is identifiable. By applying Lemma~\ref{lemma : cospark in terms of Kmin} and Lemma~\ref{lemma : cospark in terms of Kmin2},
\begin{center}
$\begin{array}{ccc}
\textrm{Cospark}(\Hbf_{\angle V}\Bbf_{\Delta})
& = & \min\limits_{\{1,2,\dots,\Gamma\}}\textrm{Cospark}(\Hbf_{\angle V}^{(\gamma)}\Bbf_{\Delta}^{(\gamma)})\\
& = & \min\limits_{\{1,2,\dots,\Gamma\}}K^{(\gamma)}\\
&=& K_{min},
\end{array}$
\end{center}
where $K_{min} = \min\limits_{1,2,\dots,\Gamma\}}K^{(\gamma)}$, $\ie,$ the smallest number of PMUs in a zone in the network. 
Since the sparsity of $\boldsymbol{\alpha}$ only takes integer values, we can rewrite the inequality and state that if, 
\begin{center}
    $\|\boldsymbol{\alpha}\|_0 \leq \ceil{\frac{K_{min}}{2} - 1}$,
\end{center}
then $\boldsymbol{\alpha}$ is identifiable. Therefore this proves the first statement in Theorem~\ref{thm : condition on sparsity of alpha - global}. 

Now we will prove the converse statement. Let $\xbf \in \mathbb{C}^N$ be an arbitrary state vector. We will prove that there exists $\boldsymbol{\alpha}$ with $\|\boldsymbol{\alpha}\|_0 = \ceil{\frac{K_{min}}{2}-1}+1 $ that is not identifiable for the state $\xbf$. In particular we explicitly construct $\boldsymbol{\alpha}$ as follows.
Without loss of generality we assume that Zone 1 in the network has the smallest number of PMUs among all zones. For legibility, let $\kappa = (\ceil{\frac{K_{min}}{2}-1}+1)$. Then we define each entry $i$ of $\boldsymbol{\alpha}^{(1)}$ as follows:
\begin{center}
$\alpha^{(1)}[i] =
\begin{cases} &
a~~~~~\textrm{if } i= 1,\dots,\kappa\\
& 0 ~~~~~\textrm{if } i=\kappa + 1, \dots, K^{(1)}
\end{cases}$    
\end{center}
where $a$ is a nonzero constant.
Furthermore, each entry of $\boldsymbol{\alpha}^{(\gamma)}$ for the rest of the zones $\gamma\in\{2,\dots,Z\}$ are set to zero, \ie:
\begin{center}
$\alpha^{(\gamma)}[i] = 0~~~~~\textrm{if } \gamma\in\{2,\dots,\Gamma\}$    
\end{center}
Let noiseless measurements $\bar{\zbf}$,
denote the measurements generated by this $\boldsymbol{\alpha}$ and an arbitrary state vector $\xbf$,~\ie,~$\bar{\zbf} = \boldsymbol{\Phi}(\boldsymbol{\alpha})\Hbf\xbf$.

In order to show that $\boldsymbol{\alpha}$ is not identifiable for x, we will prove existence of $\bar{\boldsymbol{\alpha}}\neq\boldsymbol{\alpha}$ and $\bar{\xbf}$ satisfying,
\begin{enumerate}[(i)]
    \item $\|\bar{\boldsymbol{\alpha}}\|_0 \leq \|\boldsymbol{\alpha}\|_0 $, and
    \item $
\boldsymbol{\Phi}(\boldsymbol{\alpha})\Hbf\xbf = \boldsymbol{\Phi}(\bar{\boldsymbol{\alpha}})\Hbf\bar{\xbf}
$
\end{enumerate}
We define $\boldsymbol{\bar{\alpha}}\neq \boldsymbol{\alpha}$ by only altering entries of $\boldsymbol{\alpha}$ corresponding to Zone 1. Specifically as below:
\begin{equation}
\label{eq: theorem I alpha }
\bar{\boldsymbol{\alpha}}^{(\gamma)}=
\begin{cases}
& \boldsymbol{\alpha}^{(1)} - a\cdot\boldsymbol{1}_{K_{min}}~~~~\textrm{if } \gamma=1\\ & \boldsymbol{\alpha}^{(\gamma)}~~~~~~~~~~~~~~~~~~~\textrm{if } \gamma\in\{2,\dots,\Gamma\}\\ 
\end{cases}
\end{equation}
And define state $\bar{\xbf}$ by only altering entries of $\xbf$ corresponding to Zone 1. Specifically
\begin{equation}
\label{eq: theorem I x }
\bar{\xbf}^{(\gamma)}=
\begin{cases}
& e^{j\cdot a}\cdot\xbf^{(1)}~~\textrm{if } \gamma=1\\ & \xbf^{(\gamma)}~~~~~~~~~~\textrm{if } \gamma\in\{2,\dots,\Gamma\}\\
\end{cases}
\end{equation}
Let $\bar{\bar{\zbf}}$ denote the measurements generated by $(\bar{\xbf},\bar{\boldsymbol{\alpha}})$. 
Due to the decomposibility of PMU measurements~\eqref{eq: decompose model zonewise}, the spoofed measurements in Zone $\gamma$ for $\gamma \in \{2, \ldots, \Gamma\}$, can be given as below,
\begin{equation*}
\begin{array}{ccc}
\bar{\bar{\zbf}}^{(\gamma)} & = &  \Phi_\gamma(\bar{\boldsymbol{\alpha}}^{(\gamma)})\Hbf^{(\gamma)}\bar{\xbf}^{(\gamma)}     \\
& = & \Phi_\gamma({\boldsymbol{\alpha}}^{\gamma})\Hbf^{(\gamma)}{\xbf}^{(\gamma)}
\end{array}
\end{equation*}
where last equality is due to~\eqref{eq: theorem I alpha } and~\eqref{eq: theorem I x }. Therefore,
\begin{center}
$\bar{\bar{\zbf}}^{(\gamma)} = \bar{\zbf}^{(\gamma)}$ for $\gamma\in\{2,\dots,\Gamma\}$.    
\end{center}
Now let us analyse $\bar{\bar{\zbf}}^{(1)}$ generated from attack $\bar{\boldsymbol{\alpha}}$ and state $\bar{\xbf}$. By plugging in~\eqref{eq: theorem I alpha } and~\eqref{eq: theorem I x } in~\eqref{eq: decompose model zonewise} we can write the following:
\begin{equation*}
\begin{array}{ccc}
\bar{\bar{\zbf}}^{(1)} & = &  \Phi_1(\bar{\boldsymbol{\alpha}}^{(1)})\Hbf^{(1)}\bar{\xbf}^{(1)}     \\
& = & \Phi_1({\boldsymbol{\alpha}}^{(1)} - a\cdot\boldsymbol1_{K_{min}})\Hbf^{(1)}{(e^{j\cdot a} \cdot\xbf^{(1)})}
\end{array}
\end{equation*}
Due to the fact that $\Phi_1({\boldsymbol{\alpha}}^{(1)} - a\cdot\boldsymbol1_{K_{min}})$ is a diagonal matrix with the $i^{th}$ diagonal entry equal to $e^{j(\alpha^{(1)}[i]) - a)}$, the above equation can be rewritten as below:
\begin{equation*}
\begin{array}{ccc}
\bar{\bar{\zbf}}^{(1)} & = & e^{-j\cdot a}\cdot \Phi_1({\boldsymbol{\alpha}}^{(1)})\Hbf^{(1)}{(e^{j\cdot a} \cdot\xbf}^{(1)})\\
 & = & e^{-j\cdot a}\cdot e^{j\cdot a} \cdot \Phi_1({\boldsymbol{\alpha}}^{(1)})\Hbf^{(1)}{\cdot\xbf}^{(1)}\\
& = & \Phi_1({\boldsymbol{\alpha}}^{(1)})\Hbf^{(1)}\xbf^{(1)} = \bar{\zbf}^{(1)}
\end{array}
\end{equation*}

Thus we have shown that $\bar{\bar{\zbf}} = \bar{\zbf}$
Furthermore, by the definition of $\bar{\boldsymbol{\alpha}}$;
\begin{center}
$\bar{\alpha}^{(1)}[i] =
\begin{cases} &
0~~~~~\textrm{if }i\in\mathscr\{1,\dots,\kappa\}\\
& -a ~~~~~\textrm{otherwise}
\end{cases}$    
\end{center}
and $\bar{\boldsymbol{\alpha}}^{(\gamma)} = \boldsymbol{0}$ for all $\gamma\in\{2,\dots,\Gamma\}$.
Therefore, 
\begin{center}
$\begin{array}{ccc}
\|\bar{\boldsymbol{\alpha}}\|_0 & = & K_{min} - \kappa \\
& =  &  K_{min} -  (\ceil{\frac{K_{min}}{2}-1}+1)\\
& =  &  \ceil{\frac{K_{min}-1}{2}-1}+1\\ 
& \leq & \|\boldsymbol{\alpha}\|_0\end{array}$
\end{center} 
Hence from Definition~\ref{def : global identifiability}, this attack $\boldsymbol{\alpha}$ is not identifiable. Thus we have proved that given any state $\xbf$ there exists an unidentifiable attack $\boldsymbol{\alpha}$ with sparsity level $\ceil{\frac{K_{min}}{2}-1}+1)$.
\end{proof}
\subsection{Proof of Theorem~\ref{thm : condition on sparsity of alpha - local}}
\label{proof: thm : condition on sparsity of alpha - local}
\begin{proof}
Let $\xbf$ be an arbitrary state vector $\xbf \in \mathbb{C}^N$ and $\boldsymbol{\alpha}$ be an arbitrary attack vector satisfying,
\begin{equation}
\label{eq: theorem III-2 assumption}
\|\boldsymbol{\alpha}^{(\gamma)}\|_0 \leq \ceil{\frac{K^{(\gamma)}}{2} - 1},
\end{equation}
for all $\gamma\in\{1,\dots,\Gamma\}$. We will prove that $\boldsymbol{\alpha}$ is identifiable for $\xbf$ using the proof-by-contradiction approach. 

Suppose that $\boldsymbol{\alpha}$ is unidentifiable for $\xbf$. Then, the contrapositive of Proposition~\ref{def : global identifiability} implies that there exists $\bar{\boldsymbol{\alpha}}\neq\boldsymbol{\alpha}$ such that (i) $\bar{\boldsymbol{\alpha}}\leq\boldsymbol{\alpha}$ and (ii) $\boldsymbol{\alpha} - \bar{\boldsymbol{\alpha}}\in\mathscr{R}(\Hbf_{\angle V}\Bbf_{\Delta})$. Because $\|\bar{\boldsymbol{\alpha}}\|_0 \leq \|\boldsymbol{\alpha}\|_0$ and $\boldsymbol{\alpha} \neq \bar{\boldsymbol{\alpha}}$, there exists $\bar{\gamma}\in\{1, \dots, \Gamma\}$ such that $\|\bar{\boldsymbol{\alpha}}^{(\bar{\gamma})}\|_0 \leq \|\boldsymbol{\alpha}^{(\bar{\gamma})}\|_0$ and $\bar{\boldsymbol{\alpha}}^{(\bar{\gamma})} \neq \boldsymbol{\alpha}^{(\bar{\gamma})}$. Then from the triangle inequality,
\begin{equation}
\label{eq: intermediate proof of local identifiability}
\begin{array}{ccc}
\|\boldsymbol{\alpha}^{(\bar{\gamma})} - \bar{\boldsymbol{\alpha}}^{(\bar{\gamma})}\|_0 & \leq &  \|\boldsymbol{\alpha}^{(\bar{\gamma})}\|_0 + \|\bar{\boldsymbol{\alpha}}^{(\bar{\gamma})}\|_0 
\end{array}
\end{equation}
Furthermore,
\begin{equation}
\label{eq: intermediate proof of local identifiability2}
\begin{array}{ccc}
    \|\bar{\boldsymbol{\alpha}}^{(\bar{\gamma})}\|_0 & \leq & 2\|\boldsymbol{\alpha}^{(\bar{\gamma})}\|_0 \\ & \leq &  2\ceil{\frac{K^{(\bar{\gamma})}}{2} - 1},
\end{array}
\end{equation}
where the first inequality is due to the existence of  $\bar{\gamma}\in\{1, \dots, \Gamma\}$ such that $\|\bar{\boldsymbol{\alpha}}^{(\bar{\gamma})}\|_0 \leq \|\boldsymbol{\alpha}^{(\bar{\gamma})}\|_0$ and the last inequality is due to the assumption~\eqref{eq: theorem III-2 assumption}.

Furthermore, due to the block structure of $\Hbf_{\angle V} \Bbf_\Delta$ described in~\eqref{eq: structure of H v B delta }, $\boldsymbol{\alpha} - \bar{\boldsymbol{\alpha}} \in \Rmsc(\Hbf_{\angle V} \Bbf_\Delta)$ implies that  $\boldsymbol{\alpha}^{(\bar{\gamma})}  - \bar{\boldsymbol{\alpha}}^{(\bar{\gamma})}$ is in $\Rmsc(\Hbf_{\angle V}^{(\bar{\gamma})} \Bbf_\Delta^{(\bar{\gamma})})$.
Therefore, 
\begin{center}
    $\boldsymbol{\alpha}^{(\bar{\gamma})} - \bar{\boldsymbol{\alpha}}^{(\bar{\gamma})} \in \Rmsc(\Hbf_{\angle V}^{(\bar{\gamma})}\Bbf_{\Delta}^{(\bar{\gamma})})$.
\end{center}
Since $\boldsymbol{\alpha}^{(\bar{\gamma})}$ - $\bar{\boldsymbol{\alpha}}^{(\bar{\gamma})}$ is a nonzero vector in $\Rmsc(\Hbf_{\angle V}^{(\bar{\gamma})}\Bbf_{\Delta}^{(\bar{\gamma})})$, the definition of Cospark($\Rmsc(\Hbf_{\angle V}^{(\bar{\gamma})}\Bbf_{\Delta}^{(\bar{\gamma})})$) implies that,
\begin{center}
    $\|\boldsymbol{\alpha}^{(\bar{\gamma})} - \bar{\boldsymbol{\alpha}}^{(\bar{\gamma})}\|_0 \geq \textrm{Cospark}(\Hbf_{\angle V}^{(\bar{\gamma})}\Bbf_{\Delta}^{(\bar{\gamma})})$.
\end{center}
From Lemma~\ref{lemma : cospark in terms of Kmin2} we have Cospark$(\Rmsc(\Hbf_{\angle V}^{(\bar{\gamma})}\Bbf_{\Delta}^{(\bar{\gamma})})) = K^{(\bar{\gamma})}$. Therefore, the above inequality can be rewritten as follows:
\begin{center}
    $\|\boldsymbol{\alpha}^{(\bar{\gamma})} - \bar{\boldsymbol{\alpha}}^{(\bar{\gamma})}\|_0 \geq K^{(\bar{\gamma})}$.
\end{center}
This contradicts with~\eqref{eq: intermediate proof of local identifiability2}, thereby proving the theorem statement.

Now we will prove the converse statement. Let $\xbf \in \mathbb{C}^N$ be an arbitrary state vector and $\bar{\gamma} \in \{1,..., \Gamma\}$ be an arbitrary zone. We will prove that there exists $\boldsymbol{\alpha}$ with (i) $\|\boldsymbol{\alpha}^{(\bar{\gamma})}\|_0 = \ceil{\frac{K^{(\bar{\gamma})}}{2}-1}+1$ and $\|\boldsymbol{\alpha}^{(\gamma)}\|_0 \leq\ceil{\frac{K^{(\gamma)}}{2}-1},~\textrm{for }~\gamma\in \{1,\dots,\Gamma\}\setminus{\{\bar{\gamma}\}}$, and (ii) $\boldsymbol{\alpha}$ is not identifiable for the state $\xbf$. In particular we explicitly construct such $\boldsymbol{\alpha}$ as follows.
For legibility, let $\kappa := (\ceil{\frac{K^{(\bar{\gamma})}}{2}-1}+1)$. Then we set the entries of $\boldsymbol{\alpha}^{(\bar{\gamma})}$ as follows:
\begin{center}
$\alpha^{(\bar{\gamma})}[i] =
\begin{cases} &
a~~~~~\textrm{if } i= 1,\ldots,\kappa\\
& 0 ~~~~~\textrm{if } i =\kappa + 1,\ldots,K^{(\bar{\gamma})}
\end{cases}$    
\end{center}
where $a$ is a nonzero constant.
Furthermore, for every $\gamma\in\{1,\dots,\Gamma\}\setminus\{\bar{\gamma}\}$ we set $\boldsymbol{\alpha}^{(\gamma)}$ to be an arbitrary $K^{(\gamma)}-$dimensional vector satisfying $\|\boldsymbol{\alpha}^{(\gamma)}\|_0 \leq\ceil{\frac{K^{(\gamma)}}{2}-1}$.

Let noiseless measurements $\bar{\zbf}$,
denote the measurements generated by this $\boldsymbol{\alpha}$ and an arbitrary state vector $\xbf$,~\ie,~$\bar{\zbf} = \boldsymbol{\Phi}(\boldsymbol{\alpha})\Hbf\xbf$.

In order to show that $\boldsymbol{\alpha}$ is not identifiable for x, we will prove existence of $\bar{\boldsymbol{\alpha}}\neq\boldsymbol{\alpha}$ and $\bar{\xbf}$ satisfying,
\begin{enumerate}[(i)]
    \item $\|\bar{\boldsymbol{\alpha}}\|_0 \leq \|\boldsymbol{\alpha}\|_0 $, and
    \item $
\boldsymbol{\Phi}(\boldsymbol{\alpha})\Hbf\xbf = \boldsymbol{\Phi}(\bar{\boldsymbol{\alpha}})\Hbf\bar{\xbf}
$
\end{enumerate}
We define $\boldsymbol{\bar{\alpha}} = [(\boldsymbol{\bar{\alpha}}^{(1)})^T,\ldots,(\boldsymbol{\bar{\alpha}}^{(Z)})^T]^T$ by only altering the entries of $\boldsymbol{\alpha}$ corresponding to Zone $\bar{\gamma}$ as below:
\begin{equation}
\label{eq: theorem II alpha}
\bar{\boldsymbol{\alpha}}^{(\gamma)}=
\begin{cases}
& \boldsymbol{\alpha}^{(\bar{\gamma})} - a\cdot\boldsymbol{1}_{K^{(\bar{\gamma})}}~~~~\textrm{if } \gamma=\bar{\gamma}\\ & \boldsymbol{\alpha}^{(\gamma)}~~~~~~~~~~~~~~~~~~~\textrm{if } \gamma\in\{1,\dots,\Gamma\}\setminus\{\bar{\gamma}\}\\ 
\end{cases}
\end{equation}
And define state $\bar{\xbf}$ by only altering entries of $\xbf$ corresponding to Zone $\bar{\gamma}$. Specifically,
\begin{equation}
\label{eq: theorem II x}
\bar{\xbf}^{(\gamma)}=
\begin{cases}
& e^{j\cdot a}\cdot\xbf^{(\bar{\gamma})}~~~~\textrm{if } \gamma=\bar{\gamma}\\ & \xbf^{(\gamma)}~~~~~~~~~~~~\textrm{if } \gamma\in\{1,\dots,\Gamma\}\setminus\{\bar{\gamma}\}\\ 
\end{cases}
\end{equation}
Due to the decomposibility of PMU measurements~\eqref{eq: decompose model zonewise}, the spoofed measurements in Zone $\gamma$ for $\gamma \in \{1, \ldots, \Gamma\}\setminus \{\bar{\gamma}\}$, can be given as below,
\begin{equation*}
\begin{array}{ccc}
\bar{\bar{\zbf}}^{(\gamma)} & = &  \Phi_\gamma(\bar{\boldsymbol{\alpha}}^{(\gamma)})\Hbf^{(\gamma)}\bar{\xbf}^{(\gamma)}     \\
& = & \Phi_\gamma({\boldsymbol{\alpha}}^{\gamma})\Hbf^{(\gamma)}{\xbf}^{(\gamma)}
\end{array}
\end{equation*}
where last equality is due to~\eqref{eq: theorem II alpha} and~\eqref{eq: theorem II x}. Therefore,
\begin{center}
$\bar{\bar{\zbf}}^{(\gamma)} = \bar{\zbf}^{(\gamma)}$ for $\gamma\in\{1,\dots,\Gamma\}\setminus\{\bar{\gamma}\}$. 
\end{center}
Now let us analyse $\bar{\bar{\zbf}}^{(\bar{\gamma})}$ generated from attack $\bar{\boldsymbol{\alpha}}$ and state $\bar{\xbf}$. By plugging in~\eqref{eq: theorem II alpha} and~\eqref{eq: theorem II x} in~\eqref{eq: decompose model zonewise} we can write the following:
\begin{equation*}
\begin{array}{ccc}
\bar{\bar{\zbf}}^{(\bar{\gamma})} & = &  \Phi_{(\bar{\gamma})}(\bar{\boldsymbol{\alpha}}^{(\bar{\gamma})})\Hbf^{(\bar{\gamma})}\bar{\xbf}^{(\bar{\gamma})}     \\
& = & \Phi_{\bar{\gamma}}({\boldsymbol{\alpha}}^{(\bar{\gamma})} - a\cdot\boldsymbol1_{K^{(\bar{\gamma})}})\Hbf^{(\bar{\gamma})}{(e^{j\cdot a} \cdot\xbf^{(\bar{\gamma})})}
\end{array}
\end{equation*}
Since $\Phi_{\bar{\gamma}}({\boldsymbol{\alpha}}^{(\bar{\gamma})} - a\cdot\boldsymbol1_{K^{(\bar{\gamma})}})$ is a diagonal matrix with the $i^{th}$ diagonal entry equal to $e^{j(\alpha^{(\bar{\gamma})}[i]) - a)}$, the above equation can be rewritten as below:
\begin{equation*}
\begin{array}{ccc}
\bar{\bar{\zbf}}^{(\bar{\gamma})} & = & e^{-j\cdot a}\cdot \Phi_{\bar{\gamma}}({\boldsymbol{\alpha}}^{(\bar{\gamma})})\Hbf^{(\bar{\gamma})}{(e^{j\cdot a} \cdot\xbf}^{(\bar{\gamma})})\\
 & = & e^{-j\cdot a}\cdot e^{j\cdot a} \cdot \Phi_{\bar{\gamma}}({\boldsymbol{\alpha}}^{(\bar{\gamma})})\Hbf^{(\bar{\gamma})}{\cdot\xbf}^{(\bar{\gamma})}\\
& = & \Phi_{\bar{\gamma}}({\boldsymbol{\alpha}}^{(\bar{\gamma})})\Hbf^{(\bar{\gamma})}\xbf^{(\bar{\gamma})} = \bar{\zbf}^{(\bar{\gamma})}
\end{array}
\end{equation*}
Thus we have shown that $\bar{\bar{\zbf}} = \bar{\zbf}$
Furthermore, by the definition of $\bar{\boldsymbol{\alpha}}$;
\begin{center}
$\bar{\alpha}^{(\bar{\gamma})}[i] =
\begin{cases} &
0~~~~~\textrm{if }i = 1,\dots,\kappa\\
& -a ~~~\textrm{if }i = \kappa+1, \dots, K^{(\bar{\gamma})}
\end{cases}$    
\end{center} 
Therefore, 
\begin{center}
$\begin{array}{ccc}
\|\bar{\boldsymbol{\alpha}}^{(\bar{\gamma})}\|_0 & = & K^{{(\bar{\gamma})}} - \kappa \\
& =  &  K^{{(\bar{\gamma})}} -  (\ceil{\frac{K^{{(\bar{\gamma})}}}{2}-1}+1)\\ &
\leq & (\ceil{\frac{K^{{(\bar{\gamma})}}-1}{2}-1}+1)\\ &
\leq & \|\boldsymbol{\alpha}^{(\bar{\gamma})}\|_0
\end{array}$
\end{center} 
Due to the above inequality together with the fact that $\bar{\boldsymbol{\alpha}}^{(\gamma)} = \boldsymbol{\alpha}^{(\gamma)}$ for all $\gamma\in\{1,\dots,\Gamma\}\setminus\{\bar{\gamma}\}$, implies that 
\begin{center}
$\begin{array}{ccc}
\|\bar{\boldsymbol{\alpha}}\|_0 & \leq & \|\boldsymbol{\alpha}\|_0
\end{array}$
\end{center}
Hence from Definition~\ref{def : global identifiability}, this attack $\boldsymbol{\alpha}$ is unidentifiable.
\end{proof}

\IEEEpeerreviewmaketitle
\end{document}

%% file: macros.tex
\newcommand{\ie}{\textit{i.e.}}

\newcommand{\Amsc}{\mathscr{A}}

\newcommand{\Emsc}{\mathscr{E}}

\newcommand{\Gmsc}{\mathscr{G}}

\newcommand{\Mmsc}{\mathscr{M}}
\newcommand{\Nmsc}{\mathscr{N}}

\newcommand{\Rmsc}{\mathscr{R}}

\newcommand{\Tmsc}{\mathscr{T}}

\newcommand{\Vmsc}{\mathscr{V}}

\newcommand{\mbbR}{\mathbb{R}}  

\newcommand{\Abf}{{\bf {A}}}
\newcommand{\Bbf}{{\bf {B}}}

\newcommand{\Ibf}{{\bf {I}}}

\newcommand{\Hbf}{{\bf {H}}}
\newcommand{\Pbf}{{\bf {P}}}

\newcommand{\bbf}{{\bf {b}}}

\newcommand{\ebf}{{\bf {e}}}
\newcommand{\hbf}{{\bf {h}}}

\newcommand{\rbf}{{\bf {r}}}

\newcommand{\wbf}{{\bf {w}}}
\newcommand{\xbf}{{\bf {x}}}

\newcommand{\ybf}{{\bf {y}}}
\newcommand{\zbf}{{\bf {z}}}

\newcommand{\obf}{{\boldsymbol{0}}}

\DeclarePairedDelimiter{\ceil}{\lceil}{\rceil}

\DeclareMathOperator*{\argmin}{\arg\!~\min} 
\DeclareMathOperator*{\argmax}{\arg\!~\max} 
\algnewcommand\INPUT{\item[\textbf{Input:}]}
\algnewcommand\OUTPUT{\item[\textbf{Output:}]}
\algnewcommand\INITIAL{\item[\textbf{Initialize:}]}